\documentclass{aa}

\usepackage{lscape}
\usepackage{graphicx}
\usepackage[english]{babel}  
\usepackage{graphicx}  
\usepackage[utf8]{inputenc}  
\usepackage{microtype}  
\usepackage{booktabs}  
\usepackage{rotating}  
\usepackage{lscape}
\usepackage[squaren]{SIunits}
\usepackage{natbib}
\usepackage{color}
\usepackage{amsmath}
\usepackage{amssymb}
\usepackage{soul} 

\newcommand{\water}{H$_2$O}

\newcommand{\Myr}{$M_{\sun}$\,yr$^{-1}$}
\newcommand{\mdot}{$\dot{\rm{M}}$}

\newcommand{\Lup}{$L_{\rm H_2O}^{up}$}

\newcommand{\kms}{km\,s$^{-1}$}

\newcommand{\degs}{$^{\circ}$}
\newcommand{\pad}{.\hskip-2pt$^\circ$}
\newcommand{\pam}{.\hskip-2pt$^{\prime}$}

\newcommand{\gsim}{\;\lower.6ex\hbox{$\sim$}\kern-7.75pt\raise.65ex\hbox{$>$}\;}
\newcommand{\lsim}{\;\lower.6ex\hbox{$\sim$}\kern-7.75pt\raise.65ex\hbox{$<$}\;}
\begin{document} 

\title{Water vapour masers in long-period variable stars}
\subtitle{II. The semi-regular variables \object{R\,Crt} and \object{RT\,Vir}}


\author { J.~Brand  \inst{1} 
        \and D. Engels \inst{2} 
        \and A. Winnberg \inst{3}}

\offprints{J.~Brand\\
\email{brand@ira.inaf.it}}

\institute{INAF - Istituto di Radioastronomia \& Italian ALMA Regional Centre, Via P. Gobetti 101,
           I--40129 Bologna, Italy
      \and Hamburger Sternwarte, Universit\"{a}t Hamburg, Gojenbergsweg 112,
           D--21029 Hamburg, Germany
      \and Onsala Rymdobservatorium, Observatoriev\"{a}gen,
           S--43992 Onsala, Sweden
           }

\date{Received date; accepted date}

\abstract {Water masers emitting at a radiofrequency of 22 GHz are often found in the circumstellar envelopes  of evolved stars. We monitored the \water\ maser emission of a larger sample of evolved stars of different types to study the maser properties as a function of stellar type.}
{We wish to understand the origin and evolution of the \water\ masers in circumstellar envelopes. In this paper, we take a closer look at R~Crt and RT~Vir, two nearby ($<250$~pc) semi-regular variable stars. The findings complement our monitoring results for \object{RX~Boo} and \object{SV~Peg}, two other semi-regular variable stars that we have discussed in a previous paper.}
{Within the framework of the Medicina/Effelsberg \water\ maser monitoring programme, we observed the maser emission of R~Crt and RT~Vir for more than two decades with single-dish telescopes. To get insights into the distribution of maser spots in the circumstellar envelopes at different times, to get an idea of their longevity, and, where possible, to be able to link the phenomena seen in our observations to maser locations within the envelopes, we collected interferometric data for these stars, taken within the same period, from the literature.}
{The \water\ masers in R~Crt and RT~Vir exhibit brightness variations on a variety of timescales. We confirm short-time variations of individual features on timescales of months to up to 1.5 years, as seen by previous monitoring programmes. Also decade-long variations of the general brightness level, independent from individual features, were seen in both stars. These long-term variations are attributed to brightness variations occurring independently from each other in selected velocity ranges and they are independent of the optical light curve of the stars.
Expected drifts in velocity of individual features are usually masked by the blending of other features with similar velocities. However, in RT~Vir, we found the exceptional case of a single feature with a constant velocity over 7.5 years ($<0.06$~\kms\,yr$^{-1}$).}
{We attribute the long-term brightness variations to the presence of regions with higher-than-average density in the stellar wind and hosting several clouds which emit maser radiation on short timescales. These regions typically need $\sim$20 years to cross the \water\ maser shell, where the right conditions for exciting \water\ masers are present. Different clouds contained in such a region all move within a narrow range of velocities, and so does their maser emission. This sometimes gives the impression of longer-living features in single-dish spectra, in spite of the short lifetimes of the individual components that lie at their origin, thus, naturally explaining  the longer timescales observed. The constant velocity feature (11~\kms) is likely to come from a single maser cloud, which moved through about half of RT~Vir's \water\ maser shell without changing its velocity. From this, we infer that its path was located in the outer part of the \water\ maser shell, where RT~Vir's stellar wind  has, apparently, already reached its terminal outflow velocity. This conclusion is independently corroborated by the observation that the highest \water\ maser outflow velocity in RT~Vir approaches the terminal outflow velocity, as given by OH and CO observations. This is generally not observed in other semi-regular variable stars. All four stars in our study are of optical variability type SRb, indicating the absence of periodic large-amplitude variations. Therefore, any likely responses of the maser brightness to variations of the optical emission are masked by the strong short-term maser fluctuations.}


\maketitle


\section{\label{intro} Introduction}
Water vapour (H$_2$O) maser emission of the $6_{16} - 5_{23}$ transition at 1.35-cm wavelength (22.235~GHz) is frequently found in the circumstellar shells or envelopes (CSEs) of oxygen-rich stars on the asymptotic giant branch (AGB) and in several red supergiants (RSGs). The emission consists of multiple maser lines, which are spread over a  velocity interval governed by the velocity of the stellar wind at the location of the emission region. They become detectable as soon as the mass-loss rates surpass a few times $10^{-8}$ \Myr. 
Among those with the lowest mass-loss rates are the irregular and semi-regular variable stars (SRV), of which $\approx$20\% exhibit \water\ maser emission at flux density levels $\ga500$ mJy. With stellar distances less than a few hundred parsec, maser luminosities $>10^{40}$ photons s$^{-1}$ are probed. A few of the closer SRVs have very luminous \water\ masers ($\sim 10^{43}$ photons s$^{-1}$), which made their emission lines (at times several hundred Jy-strong) preferred targets for interferometric observations to study the properties of the emission regions and of monitoring programmes for studying variability (\citealt{szymczak95}; \citealt{szymczak97}).

The early maser monitoring programmes found strong variability of the spectral shapes of the emission, of the individual contributing lines, and of the integrated flux density (e.g. \citealt{schwartz74}; \citealt{berulis83}; \citealt{likkel92}). In Mira variables and in RSGs with periodic stellar brightness variations, the  variability of the integrated \water\ maser emission was separated into two types: a variation in (delayed) response to the light variations of the central star but in sync with its period and superposed upon this is a second type attributed to irregular fluctuations of individual features in the spectra on shorter timescales, including burst events of individual maser lines lasting weeks to months (\citealt{engels88}). In SRVs of type SRb, which show only small-amplitude variability in the optical, the periodic component is absent, or at least, it is dominated by the variations of individual maser emission components (\citealt{lekht99}; \citealt{winnberg08}; see \citealt{shintani08} for the opposite result in the case of RT~Vir). 

Monitoring programmes of water masers extending over timescales longer than typical periods of variable stars ($P\sim$1-2 years) indicate the presence of a third type of variability with possible timescales of tens of years. In the case of RT~Vir, for example, \cite{mendozatorres97} reported brightness variations on timescales of between five to six years, affecting only certain velocity intervals.
Such long-term variations are difficult to bring in line with a smooth spherical wind, in which outflowing maser clouds randomly are excited and de-excited on timescales of not more than three years, corresponding to the sound-crossing times of such clouds \citep{bains03}. 

H$_2$O masers are sensitive tools that are useful for probing the physics and dynamics of the stellar wind in that part of the CSE where conditions are favourable for the excitation of these masers, that is, at radii ranging between $\sim$5 and 50~au in the case of Mira variables and SRVs  (\citealt{bowers93}; \citealt{bowers94};  \citealt{colomer00}; \citealt{bains03}; \citealt{imai03}). This \water\ maser shell is composed of clouds of sizes $\sim$2-4~au, occupying only a small fraction of the volume (filling factor $\sim$0.01), which increase their velocity by a factor of two to three  while they cross this shell \citep{bains03}. A typical wind velocity in the shell is 5--10 \kms,\ implying that the shell is crossed in $\la40$ years. The maser clouds are expected to exhibit a significant velocity gradient with typical values  of
$K_{grad} = 0.1 - 0.4$ \kms\ au$^{-1}$ \citep{richards12}, which would lead to observable velocity shifts of individual maser features in the maser profile of several \kms, if the clouds would persist that long.

In general however, the lifetimes of individual \water\ maser clouds in Mira variables and SRVs are much shorter, so that the observed velocity shifts attributed to moving clouds, as long as the maser feature is detectable, are only on the order of few tenths of a \kms. Thus, they are difficult to distinguish from shifts induced by the blending of varying maser features with velocity differences smaller than the typical line widths. In addition to short-term brightness variations, the limited lifetimes also induce considerable changes in the distribution of maser spots in interferometric maps taken many months apart. This hindered the analysis of early mapping observations (e.g. \citealt{johnston85}) and today, they place constraints on stellar distance measurements using parallaxes determined from maser spots \citep{imai19}.

\begin{table*}
\caption{Basic information on the observed semi-regular variables. 
} 
\label{centralcoords}
\begin{center}
\begin{tabular}{rlllrlccl}
\hline\noalign{\smallskip}
\multicolumn{1}{c}{Name} & 
\multicolumn{2}{c}{$\alpha$\, \, (J2000)\, \, $\delta$} & \multicolumn{1}{c}{$D^{\rm a}$} & 
 \multicolumn{1}{c}{$V_{\ast}$} & \multicolumn{1}{c}{$V_{\rm exp}$} & \multicolumn{1}{c}{$V_{\rm H_2O}$} &
 \multicolumn{1}{c}{Monitoring}
 & \multicolumn{1}{l}{Notes$^{\rm b}$} \\
\multicolumn{1}{c}{} & \multicolumn{1}{l}{\, h\,\,  m\, \,   s}
 & \multicolumn{1}{l}{\, \,  $\circ$\,\,\, $\prime$\, \,  $\prime\prime$}
& \multicolumn{1}{c}{pc} & \multicolumn{2}{c}{\kms} 
& \multicolumn{1}{c}{\kms}
 & \multicolumn{1}{c}{Period}
 & \multicolumn{1}{c}{} \\
\hline\noalign{\smallskip}
R~Crt  &11:00:33.9&$-$18:19:29& 236$\pm$12 & 11.3 &11.7 & 1.6--21.8 & 1990-2011  & 1,2,4 \\
RT~Vir  & 13:02:37.9 & +05:11:09 & 226$\pm$7 & 17.3 &9 &  7.4--27.0 & 1987-2011 & 1,2,3,4 \\
\noalign{\smallskip}
\hline
\end{tabular}
\end{center}
$a:$\ References for distances: R~Crt: \cite{gaiacoll18}; RT~Vir: \cite{zhang17}\\
$b:$\ References for stellar- and expansion velocities: 
1.~\cite{cernicharo97}; 2.~\cite{kerschbaum99}; 3.~\cite{lekht99}; 
4.~\cite{diazluis19}\\
%

\end{table*}

In order to improve our understanding of the properties of \water\ maser variability for different types of evolved stars, we started the Medicina/Effelsberg \water\ maser monitoring programme, comprising, in total, observations taken between 1987 and 2015. In \cite{winnberg08} (hereafter, Paper I), we presented the results for two SRVs RX~Boo and SV~Peg. The \water\ maser emission of RX~Boo was found during 1990-1992 in an incomplete ring with an inner radius of 15~au and a shell thickness of 22~au, which is only partially filled. The variability of \water\ masers in RX~Boo and SV~Peg is due to the emergence and disappearance of maser clouds with lifetimes of $\sim$1~year. The emission regions do not fill the shell of RX~Boo evenly; this asymmetry in the spatial distribution persists at least one order of magnitude longer than the cloud lifetimes.

In this paper, we present the single-dish data for approximately 2 decades of monitoring for R~Crt and RT~Vir, two other SRVs of type SRb. We use data from other single-dish monitoring programmes made before and in parallel to our project, as well as interferometric maps available from the literature, to derive their \water\ maser properties, with emphasis on the long-term variations. Basic information on the two stars is given in Table~\ref{centralcoords}. 
The table gives the name of the object in column (col.)~1; in cols.~2 and 3, we give the reference coordinates of the stars; in col.~4, we give the distance to the stars, the sources for which are given in the footnote. All linear sizes given in this paper are scaled to these stellar distances. The stellar radial velocity, $V_{\ast},$ and the final expansion velocity in the CSE, $V_{\rm exp}$, are shown in cols.~5 and 6, respectively. In col.~7, we give the range in velocity over which \water\ emission was found during the monitoring period, given in col.~8., which indicates the period over which we have a more or less regular coverage with observations. In the last column (Notes), we identify the references used for the stellar and expansion velocities, listed in the footnote. The adopted values for these velocities are estimates of the central velocities ($V_{\ast}$) and the half-widths ($V_{\rm exp}$) of the emission profiles of various CO-transitions and of the 1612 MHz OH-transition reported in the cited papers, which almost always agree to within $\pm 1$~\kms.

In forthcoming papers, we will present the results from our monitoring programme of the Mira variables (Winnberg et al., in preparation) and Red Supergiants (Brand et al., in preparation). The results for OH/IR stars will follow. Monitoring results of one of them, OH~39.7+1.5, were presented by \cite{engels97}. The current paper describes the radio observations (Sect. \ref{observations}), our analytical methods (Sect. \ref{presdata}), and the \water\ maser properties of R~Crt (Sect. \ref{rcrt-main}) and RT~Vir (Sect. \ref{rtvir-main}). It concludes with a discussion on maser brightness variability properties, maser luminosities, movements in the \water\ maser shell, and constraints on the standard model for CSEs with respect to SRVs (Sect. \ref{discussion}), and finishes with a summary of the main results (Sect. \ref{conclusions}).

\section{Observations \label{observations}}
Single-dish observations of the H$_2$O($6_{16}-5_{23}$) maser line at 22235.0798~MHz were made with the Medicina 32-m and Effelsberg 100-m telescopes at typical intervals of a few months. Initial observations began in 1987 with the Medicina telescope and regular monitoring was performed between 1990 and 2011. Some additional spectra were taken in 2015. For both stars, one spectrum taken between 1987 and 1989 was already published by \cite{comoretto90}. The Effelsberg telescope participated in the monitoring programme between 1990 and 1999. 
Spectra taken in December 1992 and April 1993 were previously published by \cite{szymczak95}.

\subsection{Medicina}
Between March 1987 and October 2015, we searched for H$_2$O($6_{16}-5_{23}$) maser emission with the Medicina 32-m telescope\footnote{The Medicina 32--m VLBI radiotelescope is operated by INAF--Istituto di Radioastronomia.} towards R~Crt and RT~Vir  (Table~\ref{centralcoords}). We used a bandwidth of 10~MHz and 1024 channels, resulting in a resolution of 9.76~kHz (0.132~\kms); the half-power beam width  (HPBW) at 22~GHz was $\sim$1\pam 9. During this period, the stars were observed four or five times per year in separate sessions. For more information on the changes in the system during these years, see Paper I.

The telescope pointing model was typically updated a few times per year, and was quickly checked every few weeks by observing strong maser sources (e.g. W3~OH, Orion-KL, W49~N, Sgr~B2, and W51). The pointing accuracy was always better than 25\arcsec\ and the rms residuals from the pointing model were on the order of 8\arcsec--10\arcsec. 

The observations were taken in total power mode, with both ON and OFF scans of a 5~min duration. The OFF position was taken 1\pad 25 E of the source position to rescan the same path as the ON scan. Usually two ON-OFF pairs were taken at each position. Only the left-hand circular (LHC) polarisation output from the receiver was registered\footnote{In Paper I this was erroneously reported as only RHC (right-hand circular).}.

The observations were embedded in a larger programme. We could thus determine the antenna gain as a function of elevation by observing, several times during the day, the continuum source DR~21 -- for which we assume a flux density of 16.4 Jy after scaling the value of 17.04~Jy given by \cite{ott94} for the ratio of the source size to the Medicina beam -- at a range of elevations. Antenna temperatures were derived from total power measurements in position switching mode. The integration time at each position was 10 sec with 400~MHz bandwidth. 

The daily gain curve was determined by fitting a polynomial curve to the DR~21 data; this was then used to convert antenna temperature to flux density for all spectra taken that day. From the dispersion of the single measurements around the curve, we found the typical calibration uncertainty to be 20\%.

\subsection{Effelsberg}
Between 1990 and 1999, we also observed the sources with the Effelsberg 100-m telescope\footnote{The Effelsberg 100-m radiotelescope is operated by the Max-Planck-Institute f\"ur Radioastronomie, Bonn}.
We used 18--26~GHz receivers with cooled masers as pre-amplifiers to observe the $6_{16}\rightarrow 5_{23}$ transition of the water molecule.
It was only the LHC polarisation direction that was recorded, as circumstellar water masers were found to be non-polarised to limits of a few percent (\citealt{barvainis89}). At 1.3~cm wavelength the beam width is $\sim$40\arcsec\ (HPBW).  We observed in total power mode integrating ON and OFF the source for typically 5--10~min each. At `ON-source' the telescope was positioned on the coordinates given in Table \ref{centralcoords}, while the `OFF-source' position was displaced 3\arcmin\ to the east of the source. 

The backend consisted of a 1024 channel autocorrelator. Observations were made with a bandwidth of 6.25~MHz, centred approximately on the stellar radial velocity, $V_{\ast}$. The velocity coverage was $70-80$~\kms\ and the velocity resolution was 0.08~\kms. 
For procedures to reduce the spectra and the calibration, we refer to Paper I. We estimate that the flux density values are not reliable to better than 20--30~\%.

\section{Presentation of the data \label{presdata}}
In the following sections, we present and discuss the data on the stars in our sample. Here, we describe the tools and define the parameters we used for this purpose. All maser spectra for the stars are presented in the Appendix\footnote{Spectra will be available at the CDS via anonymous ftp to cdsarc.u-strasbg.fr (139.79.128.5) or via http://cdsweb.u-strasbg.fr/cgi-bin/qcat?J/A+A/}. For each star, we also show a selection of the spectra taken over the years, presented in the sub-sections where they are mentioned (see Fig. \ref{fig:rcrt_sel} for an example). 

\subsection{Diagnostic plots}
\begin{figure}
\includegraphics
[width=\columnwidth]
{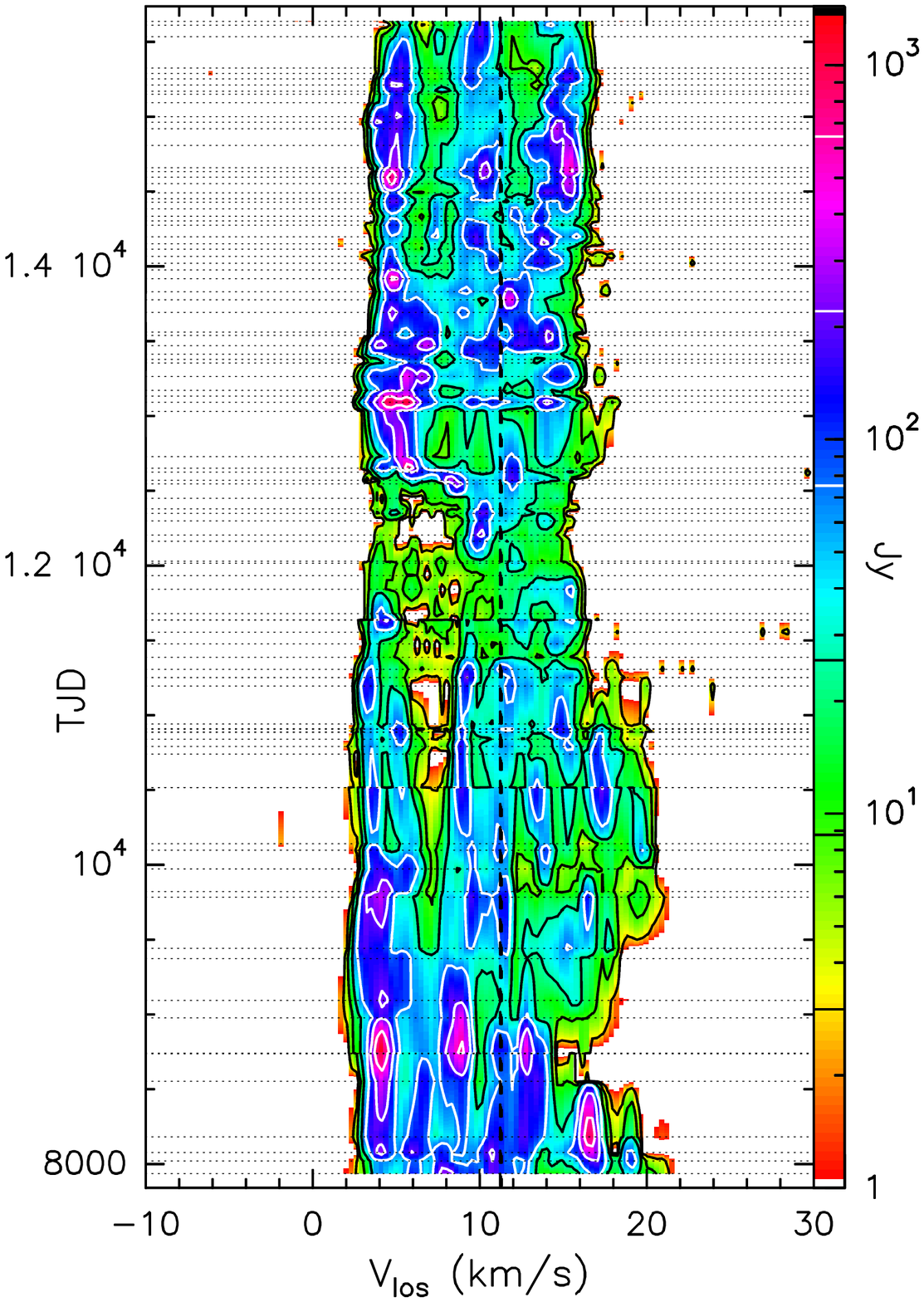}
\caption{Flux density versus line-of-sight velocity, $V_{\rm los}$, as a function of time (FVt)-plot for R~Crt. Each horizontal dotted line indicates an observation (spectra taken within 4 days from each other were averaged). Data are resampled to a resolution of 0.3~\kms\ and only emission at levels $\geq 3\sigma$ is shown. The first spectrum in this plot was taken on 17/2/90; JD = 2447939.5, TJD = 7939. The last spectrum is from 20/3/2011. The vertical dashed line marks the stellar velocity $V_{\ast}$ (Table~\ref{centralcoords}). The flux density scale is shown by the bar on the right; the black and white horizontal lines mark the values of the drawn contours.}
\label{fig:rcrt-fvt}
\end{figure}

\noindent
For each star, we show a number of plots that summarise the behaviour of the water maser emission in time, intensity, and velocity-range: 
\smallskip
\hfill\break\noindent
\underline{\it FVt-plot:}\ The time variation of the maser emission is visualised by plotting the flux density versus time and line-of-sight velocity, $V_{\rm los}$, in a so-called FVt-diagram (cf. \citealt{felli07}). An example is shown in Fig.~\ref{fig:rcrt-fvt}. Each horizontal dotted line indicates an observation (spectra taken within four days from each other were averaged). Between consecutive observations linear interpolation was applied; when there is a long time-interval between two consecutive observations, this produces an apparent persistence or increase in the lifetime of a feature. The data were resampled to a resolution of 0.3~\kms\ and only emission at levels $\geq 3\sigma$ is shown. Time is expressed in truncated Julian date format (JD$-2440000.5$)\footnote{Truncated Julian Date, which differs from MJD, the modified JD, in that an extra 40000 days are subtracted.}; correspondence with civilian calendar dates is given in the figure captions. 
While for all sources, we also took four spectra in 2015, the last spectra used in the FVt-plots are from March 2011 in order to avoid a four-year gap. 
\hfill\break\noindent
\underline{\it Upper envelope spectrum:}\ Obtained by assigning to each velocity channel the maximum (if  ${>} 3\sigma$) signal detected during our observations (including spectra taken before and after the formal 'monitoring period', as defined in Table~\ref{centralcoords}). This 'envelope' represents the maser spectrum if all velocity components were to emit at their maximum level and at the same time; the spectra are resampled to a resolution of 0.3~\kms\ first. See Fig.~\ref{fig:rcrt-upenv} for an example. 
\hfill\break\noindent
\underline{\it Lower envelope spectrum:}\ Same as the upper envelope, but obtained by finding the minimum flux density in each velocity channel, setting it to zero, unless it is ${>} 3\sigma$. When the lower envelope spectrum is zero over the whole velocity interval, it indicates that no velocity component is constantly present above a 3$\sigma$ level during the observing period. This does not necessarily imply that the maser has been quiescent at all velocities at some point in time, because the emission may occur at different velocities at different times. Also here, spectra were resampled to a resolution of 0.3~\kms\ first. An example is shown in Fig.~\ref{fig:rcrt-loenv}. Because the presence of a single noisy spectrum in the sample may cause weaker permanently present emission components to not be visible, the lower envelope spectrum should be interpreted with some caution.
\hfill\break\noindent
\underline{\it Detection-rate histogram:}\ Shows the rate-of-occurrence of maser emission above the 3$\sigma$ noise level for each velocity channel, both in absolute numbers (left axis) as in percentage (right axis). This simply counts, for each channel, the number of times the flux density in the channel is greater than the 3$\sigma$ noise level of the spectrum. Also, in this case, all the observations were used and spectra were resampled to a resolution of 0.3~\kms\ first. An example is shown in Fig.~\ref{fig:rcrt-histo}. 
\hfill\break\noindent
\underline{\it Radio (maser) light curves:}\ Obtained by plotting integrated flux densities versus TJD. 
The integrated flux density $S(tot)$ is determined over a fixed velocity interval encompassing all velocities at which maser emission was detected at least once during the monitoring period. The velocity intervals chosen enclose the range of velocities with maser emission shown in the upper envelope plots and detection-rate histograms. This choice is superior to the use of peak flux densities of individual peaks as they are not necessarily representative for the general strength since the maser profile varies over time. 
\hfill\break\noindent

\subsection{Velocities and velocity ranges}
\noindent
We analyse the observed velocity ranges of the \water\ maser emission in the framework of the standard model for CSEs in evolved stars \citep{hoefner18}. The model assumes that the stars have radially symmetric outflowing winds, which form a spherical shell of dust and gas around them. 
The winds are accelerated so that the outflow velocity, $V_{\rm out}$, is increasing with radial distance from the star before it reaches the final expansion velocity, $V_{\rm exp}$. In the shells, different molecules emit line radiation at different distances from the star, with the regions of SiO, \water,\ and OH maser emission and of CO thermal emission from the lower rotational transitions (J=1--0 and J=2--1) forming a sequence of increasing distances.

The velocity range over which \water\ maser emission can be expected is, therefore, well-constrained by the velocity ranges given by the OH maser and CO thermal emission. Both species are found beyond the typical \water\ maser shells in regions where the wind acceleration has already ceased and the outflow velocity is constant \citep{hoefner18}. In general, the OH and CO velocity ranges coincide, so that the outflow velocities derived from them are considered as final expansion velocities. 
Based on the standard model, the \water\ maser outflow velocities, $V_{\rm out},$ should obey the rule, $V_{\rm out} \le V_{\rm exp}$. 
Thus, the observed \water\ maser velocities $V_{\rm los}$  are expected to be in the range of $V_{\ast}-V_{\rm exp} \le  V_{\rm los} \le V_{\ast}+V_{\rm exp}$.
\hfill\break\noindent

We use the detection-rate histogram for the determination of the observed maximum extent of the \water\ maser velocity range $\Delta V_{\rm los}$ (hereafter, 'maximum velocity range') valid for the period of observations. Here, $\Delta V_{\rm los} = V_r-V_b$ is measured from the velocities at the blue ($V_{\rm b}$) and red ($V_{\rm r}$) extremes by conservatively requiring presence of at least five detections at the velocity limits. This choice discards individual faint emission peaks scattered in the histogram that are well outside the bulk emission and which are visible also in the FVt-plots. A random inspection of the spectra shows that these are very likely noise peaks and not actual maser components. This method gives accurate values ($\approx 0.15$ \kms) for the maximum velocity range, $\Delta V_{\rm los}$.

We estimate the maximum outflow velocity, $V_{\rm out}^{\rm max},$ reached in the \water\ maser shell from $V_{\rm b}$ and $V_{\rm r}$ using $V_{\rm out}^{\rm max} = max\{V_{\ast}-V_{\rm b}, V_{\rm r}-V_{\ast}\} - 0.5 \times \delta V_{\rm w}$, where $\delta V_{\rm w}$ is the typical line width close to the base of the maser spectral features. Here,
$\delta V_w$ acts as correction factor for the broadening of the detection-rate histogram due to the emission determining its edges. We measured, in both stars, the full width at half maximum (FWHM) of the Gaussian fit for selected maser spectral features without evidence for blending, and found $FWHM = 1.0\pm0.3$ \kms. For the line widths close to the base of the features, we adopted $\delta V_{\rm w} = 1.5 \times FWHM = 1.5$ \kms. 
In the case of spherical symmetry, we expect that the centre of the \water\ maser velocity range is $V_{\rm c} =  (V_{\rm b}+V_{\rm r})/2 = V_{\ast}$.

We should note that the velocity range of individual observations and the maximum velocity range do vary with time because of two effects. First, for certain periods of time, the outermost features may fall in brightness below the detection limit, leading to an apparent variation of the observed velocity range. And second, maser emission might be excited out to larger or smaller distances for periods of time leading to a real increase or decrease of the maximum velocity range, respectively. 

\noindent

\begin{figure*}
\resizebox{18cm}{!}{
\includegraphics{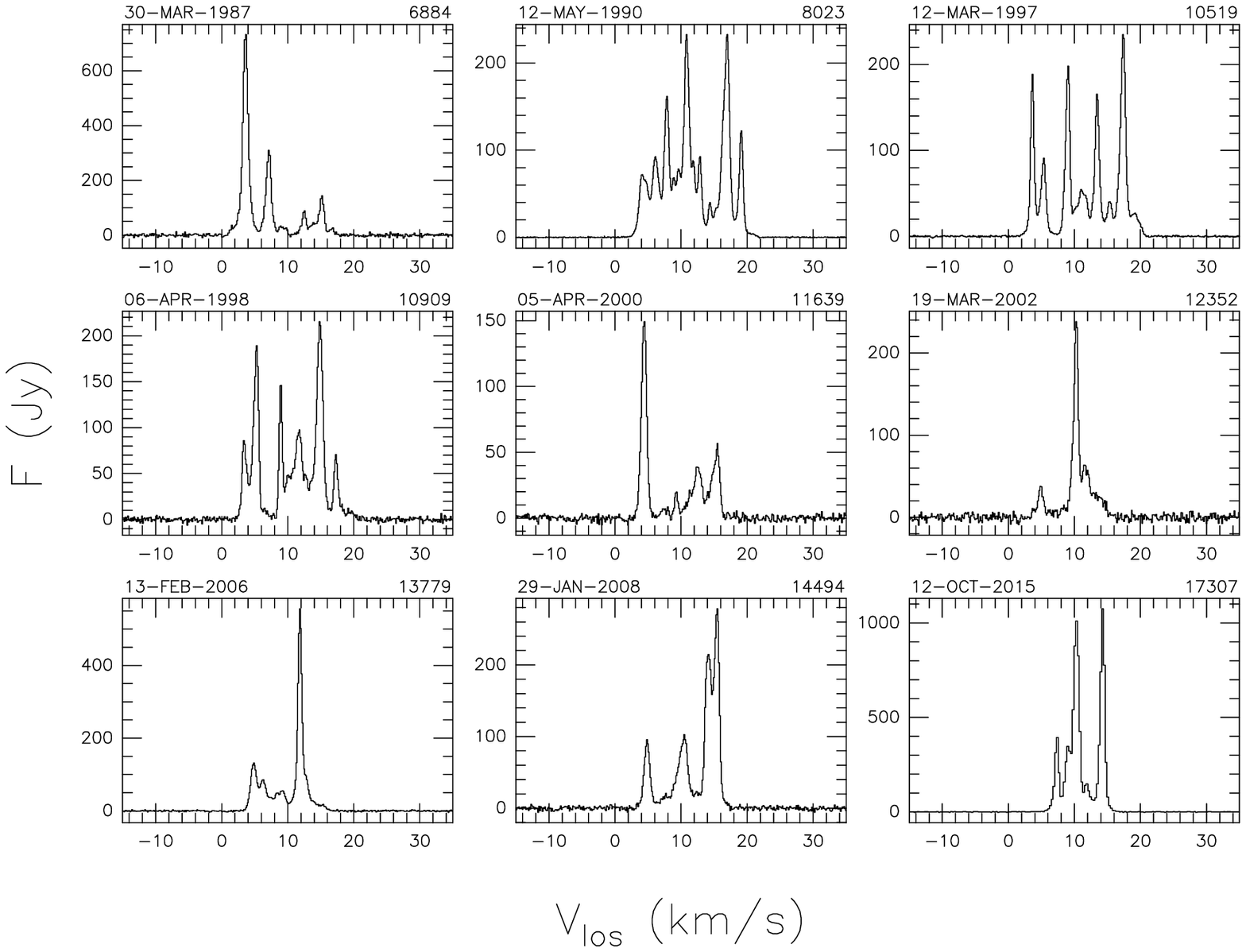}}
\caption{Selected H$_2$O maser spectra of R~Crt. The calendar date of the observation is indicated on the top left above each panel, the TJD (JD-2440000.5), on the top right.}
\label{fig:rcrt_sel}
\end{figure*}

\section{R~Crt \label{rcrt-main}}
R~Crt is a semi-regular variable AGB star at a distance of $236\pm12$ pc \citep{gaiacoll18}. We adopt as radial velocity of the star $V_{\ast} = 11.3\pm0.1$ \kms\ and as a final expansion velocity $V_{\rm exp} = 11.7\pm0.3$ \kms, as determined from circumstellar CO emission (see Table~\ref{centralcoords}).

The \water\ maser emission from R~Crt was first detected in 1971 and followed up in the following months by \cite{dickinson73}. In these observations non-detections had an upper limit to the peak flux density of 15~Jy. They found strong features with flux densities $F_\nu > 100$~Jy in the velocity range $9 < V_{\rm los} < 19$ \kms. The frequent observations of this strong maser carried out until 1985 were discussed by \cite{engels88}. Additional emission was detected since the mid-1970s at velocities $0 < V_{\rm los} < 9$ \kms\ \citep{gomezbalboa86}. The maser features were varying strongly and uncorrelated, causing the shape of the entire emission profile to vary strongly as well. Year-long phases with rather weak maser emission with flux densities $<$20~Jy, as in the mid-1970s, were alternated with bright phases with strong maser features, as in 1984/1985, when peak flux densities of up to 600~Jy were observed. No interferometric observations were made before the start of our observations.

\subsection{Results from the  \water\ maser monitoring programme}
\subsubsection{Profile characteristics and velocity ranges  \label{sec:rcrt-velo-profile}}
Representative maser spectra of R~Crt from our observations between 1987 and 2015 are shown in Fig.~\ref{fig:rcrt_sel}, while all spectra are collected in Fig.~\ref{fig:rcrt_all} (Appendix). An overview on the general behaviour of the maser variations is shown in the FVt plot (Fig.~\ref{fig:rcrt-fvt}). The emission pattern is complex, seemingly without any particular velocity intervals dominating for more than a couple of years consecutively (but see Fig.~\ref{fig:rcrt-lcurve}). Typically, the strongest feature had a flux density of $>$100 Jy, but on several occasions, even $1000-1500$~Jy was reached. Velocity intervals where bright maser features are present more often than in others: $V_{\rm los} \approx$ 3-6, $\approx$ 7-11, and $\approx$ 14-18~\kms. 
The emission in the velocity region around $\sim$5 \kms\ was also bright during the monitoring observations of \cite{gomezbalboa86} between 1976 and 1981 and of \cite{berulis83} between 1980 and 1982. 

\begin{figure}
\resizebox{9cm}{!}{\rotatebox{270}{
\includegraphics
{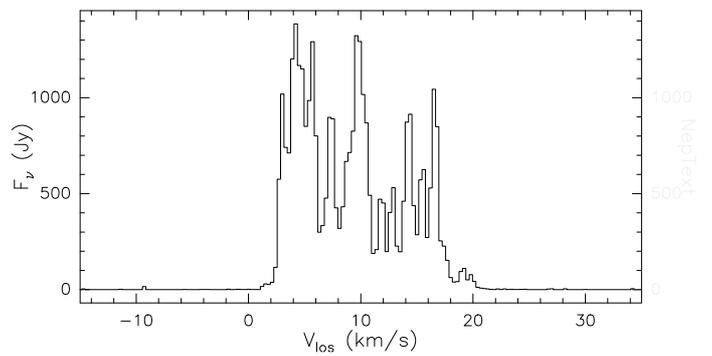}}}
\caption{Upper envelope spectrum for R~Crt; 1987-2015. This would be the maser spectrum if all velocity components were to emit at their maximum level (but $> 3\sigma$, in channels resampled to a 0.3~\kms\ resolution) at the same time.}
\label{fig:rcrt-upenv}
\end{figure}

\begin{figure}
\resizebox{9cm}{!}{\rotatebox{270}{
\includegraphics
{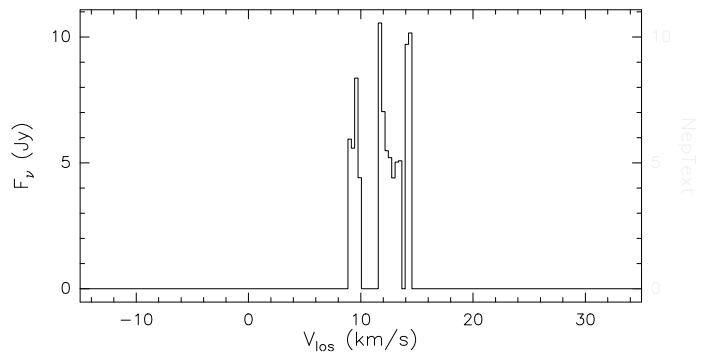}}}
\caption{Lower envelope spectrum for R~Crt; 1987-2015. This plot shows the minimum flux density level in each 0.3~\kms\ channel during the monitoring period. The flux density was set to zero if $< 3\sigma$.}
\label{fig:rcrt-loenv}
\end{figure}

\begin{figure}
\resizebox{9cm}{!}{\rotatebox{270}{
\includegraphics
{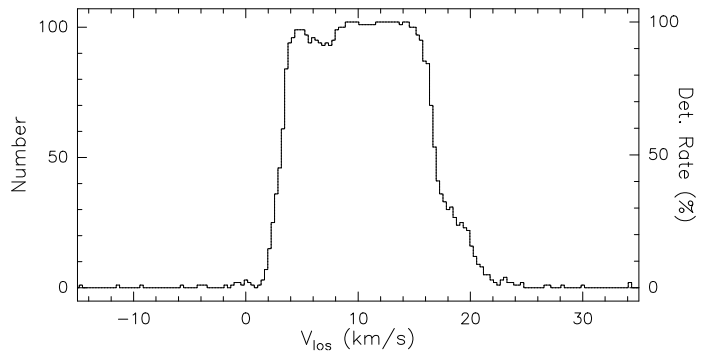}}}
\caption{Detection-rate histogram for R~Crt; 1987-2015. This figure shows how often the emission in a 0.3~\kms\ -wide channel reached a flux density $> 3\sigma$ during the monitoring period.}
\label{fig:rcrt-histo}
\end{figure}

The upper envelope spectrum (Fig.~\ref{fig:rcrt-upenv}) and the detection-rate histogram (Fig.~\ref{fig:rcrt-histo}) show that emission was detected between $1.6 < V_{\rm los} < 21.8$ \kms\ ($\Delta V_{\rm los} = 20.2$ \kms) 
for flux densities $>3\sigma$, which encompasses all the maser feature velocities ever seen, except for weak emission at velocities $V_{\rm los} < 1.4$ \kms\ detected by \cite{bowers94} in 1988. This emission is also in conflict with the final expansion velocity of R~Crt's stellar wind (Table \ref{centralcoords}) and has never been confirmed by subsequent observations. Only in the central velocity interval, 8--15 \kms,\ emission was always detected (lower envelope spectrum; Fig.~\ref{fig:rcrt-loenv}). The maximum outflow velocity in the \water\ maser shell is found to be $V_{\rm out}^{\rm max} = 10.0\pm0.5$ \kms. 
\water\ maser emission is therefore emitted over $\sim$85\% of the complete velocity range $V_{\ast} \pm V_{\rm exp}$, although the emission was not detectable at all velocities all the time. The maximum velocity range $\Delta V_{\rm los}$ is centred on the stellar velocity \mbox{($\mid V_{\rm c}-V_{\ast}\mid = 0.4$ \kms)}, as expected from spherical symmetry of the \water\ maser shell.
The detection-rate histogram shows that emission between 4 and 16 \kms\ occurs with equal frequency, while emission at $V_{\rm los} < 3$ \kms\ is seen only occasionally and at $V_{\rm los} > 17$ \kms\ in brief periods only ($<1991$ and $1994-1998$; see FVt-plot). In general, the variability pattern of the \water\ maser of R~Crt between 1987 and 2015 is similar to the pattern reported for the previous years. 

The \water\ maser velocity range shows remarkable variations, which is mainly caused by the only occasional detection of emission at $>18$ \kms\ above the 3$\sigma$ threshold used in the FVt-plot (Fig. \ref{fig:rcrt-fvt}). 
These variations are purely consequences of brightness variations, where, in particular, in the red part of the profile, the emission is weak for extended periods of time. The detected velocity range in individual observations of R~Crt is therefore strongly dependent on its maser brightness, especially at the most blue- and redshifted velocities.

\subsubsection{The \water\ maser light curve 1987 -- 2011 \label{rcrt_lcurves} }
\begin{figure}
\resizebox{9cm}{!}{\rotatebox{0}{
\includegraphics
{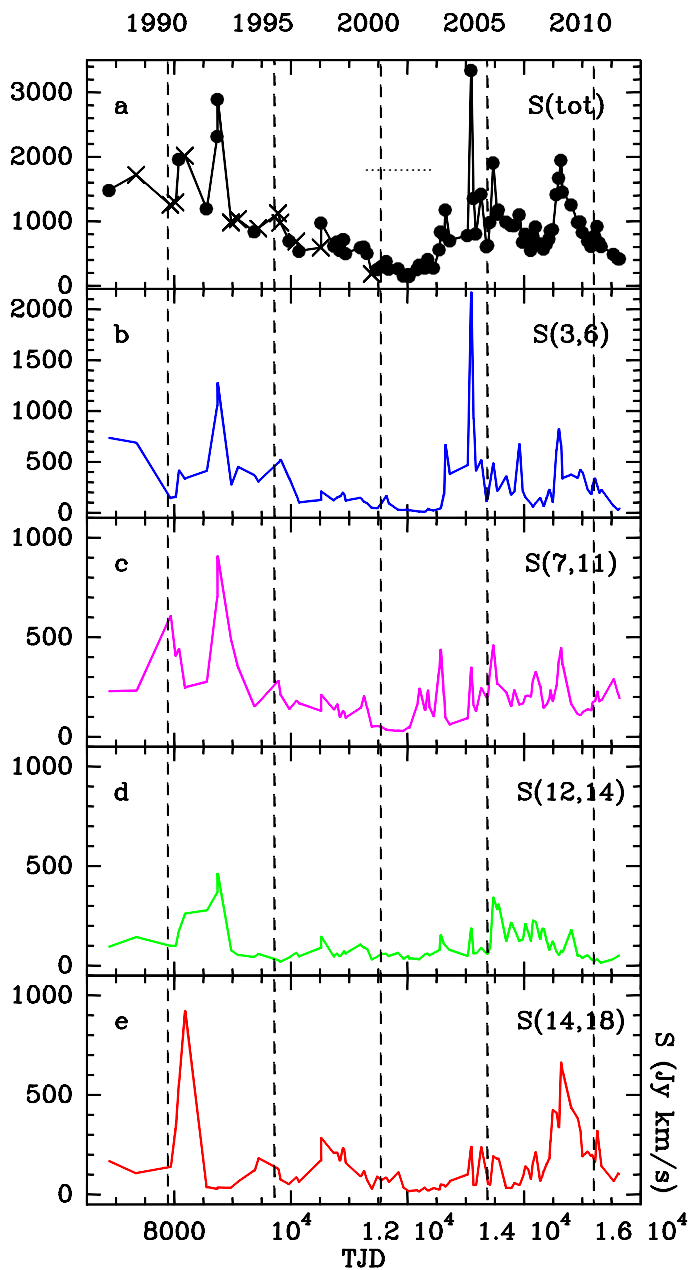}}}
\caption{\water\ maser light curves of R~Crt in 1987-2011, showing the integrated flux density, $S$, in different velocity intervals versus TJD (= Julian Date - 2440000.5). Spectra taken within three days were averaged before determining $S$. The vertical lines connect the first day of each of the years indicated at the top of the figure with the corresponding TJD on the lower axis. {\bf a.}\ $S(tot)$ is the total flux of all maser emission, between $0 < V_{\rm los} < 22$ \kms. Observations marked with crosses are from Effelsberg, the dots mark data from Medicina. 
A phase of relatively weak maser activity in 5/1999 - 5/2002 is marked by a horizontal dotted line.
{\bf b.--e.}
$S$ is the integrated flux in the velocity intervals given in the individual panels.}
\label{fig:rcrt-lcurve}
\end{figure}

Figure \ref{fig:rcrt-lcurve} shows in the upper panel the radio light curve of R~Crt of the \water\ maser emission, $S(tot),$ obtained by determining the integrated flux density in the velocity interval $0 < V_{\rm los} < 22$~\kms. 
$S(tot)$ had, on average, a level of 780~Jy \kms, but showed significant long-term  variations. Between 1987 and 1999 (TJD $\approx 6500-11300$), the integrated flux density decreased by a factor of $\approx3$, went through an extended minimum lasting about two years with a level of $\approx$200~Jy \kms\ and less, and recovered from the end of 2002. Until 2011, the flux stayed close to the average level. In 2015, the maser was found in a bright stage with an integrated flux density between 2000 and 3300~Jy \kms\ (not shown in the FVt-plot and the radio light curve). The high flux was due to three maser features at 7.5, 10, and 14.5~\kms, which (independently of each other) reached flux densities $>$1000~Jy during the year (Fig.~\ref{fig:rcrt_sel} and Appendix, Fig.~\ref{fig:rcrt_all}).

The other panels of Fig. \ref{fig:rcrt-lcurve} show light curves restricted to the integrated flux density, $S$, in the velocity intervals, $V_{\rm los}$=3-6, 7-11, 12-14, and 14-18~\kms. The first two velocity intervals and the last one temporarily contained bright maser features ($>500$ Jy), while the brightness variations in the third interval (12-14~\kms) were more moderate. The outstanding bright peaks in the velocity-dependent light curves are due to individual maser bursts, which were strong enough  to dominate even the overall integrated light curve.  The first very prominent burst occurred in the 14-18~\kms\ velocity interval with a peak in Oct. 1990 (TJD$\sim$8200) caused by a feature at $V_{\rm los}$=16.5 \kms. Shortly afterwards, in May 1992 (TJD$\sim$8740), a burst occurred in the blue velocity interval $V_{\rm los}$=3-6 at $V_{\rm los}\sim$4.0 \kms,\ accompanied by a weaker burst in the neighbouring 7-11~\kms\ interval at $V_{\rm los}$=9.5~\kms. The 9.5~\kms\ burst dominated the spectral profile later in the year (Dec. 1992), while the $V_{\rm los} \sim$4.0~\kms\ feature had decreased already below the 200~Jy level.

After the eight year-long phase of declining brightness  (1995--2002) new bursts developed in 2002--2004 in the $V_{\rm los}$=3-6~\kms\ velocity interval, reaching maximum brightness in Jan. 2003 (TJD$\sim$12650) at $V_{\rm los}$=5.8~\kms\ and 14 months later (Mar. 2004; TJD$\sim$13100) at $V_{\rm los}$=4.0 and 5.8~\kms,\ simultaneously. In 2008 the integrated light curve $S(tot)$ reached another peak, which was generated by a superposition of three bursts at velocities $V_{\rm los}$=4.7, 10.2 and 15.4~\kms,\ in all three velocity intervals, which had already shown bursts previously. They reached their maximum values non-simultaneously between May and July (TJD$\sim$14630$\pm$30). 

Although we singled out only a few spectacular events here, bursts with brightness increases by factors of between about two and five are frequent. The duration of the bursts is $\sim$0.5-1.5 years. The fluctuations have a range of amplitudes and the bursts are probably the fluctuations with the largest ones. Inversely, the absolute minimum seen in Fig.~\ref{fig:rcrt-lcurve}, which is around 2000-2002 (TJD$\sim$12000) at the end of the general decline of the emission, is also a consequence of absence of stronger fluctuations over the whole velocity range. 

The integrated maser light curve as well as the velocity-dependent ones are therefore a superposition of long-term variations (timescales: many years) of the brightness level and short-term fluctuations (timescales: months or possibly even shorter).
The long-term variations are exemplified by the $V_{\rm los}$= 3-6~\kms\ emission, which has been seen prominently since 1976 (\citealt{gomezbalboa86}; \citealt{berulis83}), and which dominates the maser profile over most of the time except between 1996 and 2002, encompassing the phase when the maser emission was weak in general in R~Crt. 

The brightness fluctuations of individual maser features are accompanied by apparent shifts in velocity. However, in general real shifts cannot be disentangled from shifts caused by independently varying lines, which have velocity separations smaller than individual line widths. 

It is known that \water\ masers in Mira variables respond to brightness variations of the central stars (\citealt{schwartz74}; \citealt{rudnitskii05} and references therein; Winnberg et al., in preparation). It is therefore possible that the strong irregular maser bursts are triggered by events, which can be identified also in the optical light curve of R~Crt. The optical light curve (V-band) was obtained from AAVSO (Kafka, 2018\footnote{Observations from the AAVSO International Database, https://www.aavso.org}) for the years 1987--2015 encompassing the monitoring programme. Unfortunately, the AAVSO light curve is only sparsely sampled, and shows no irregular brightenings that could be linked to the observed maser bursts. We performed a periodogram analysis of the optical as well as of the radio light curve, finding no evidence for periodicity. 

\subsection{Comparison with the single-dish VERA-Iriki and Kagoshima monitoring programmes}
Between September 2003 and November 2006, the maser of R~Crt was observed by the VERA-Iriki monitoring programme (\citealt{shintani08}) with revisits every month. With a sensitivity of $\approx$1 Jy, similar to ours, they detected emission between 3 and 17 \kms. This is a smaller velocity range than what we observed in 1994--1998 (TJD $\sim$9500-11000), but it matches our velocity range for the time in common (TJD $\sim$13000-14000; cf. Fig.~\ref{fig:rcrt-fvt}). They concluded that the lifetime of most of the individual maser features is relatively short ($\la$300 days). 
Our lower envelope spectrum (Fig.~\ref{fig:rcrt-loenv}) suggests that there are several velocity channels that show emission over the entire period of monitoring, that is, about two and a half decades. Because of the possible presence of velocity shifts, insufficient spectral resolution, and overlapping components, it is unlikely that these represent individual maser features coming from clouds living that long. 

Another monitoring programme of the \water\ maser, with revisits every few days, was carried out by \cite{sudou17} between January 2008 and March 2009 (TJD $\sim$14500 - $\sim$14900) with the Kagoshima 6-m radio telescope. They found velocity shifts of up to 0.9 \kms\ for the strongest features at 5, 10.5, and 15.5~\kms\ permanently visible during this period (cf. the 29 January 2008 spectrum in Fig.~\ref{fig:rcrt_sel}). The shifts were considered to be due to acceleration and deceleration of the emitting gas in the CSE. A detailed analysis of our spectra, however, shows that all three features are blends of maser lines with slightly different velocities. Such lines, also found by Sudou et al. also at velocities away from the strong features, are often sufficiently bright enough to be distinguished only for a matter of weeks. A secular intensity variation of maser lines close in velocity space on timescales of weeks to months naturally explains the velocity variations seen in these strong features over timescales of a year. 

Over the course of our monitoring programme, a few additional \water\ maser spectra with profiles consistent with ours were recorded in 1991 by \cite{takaba94, takaba01} and in 2009 by \cite{kim10}.

\subsection{Interferometric observations since 1987 \label{since87}}
The first interferometric observations were obtained with the VLA in 1988 and 1990 by \cite{bowers94} and \cite{colomer00}, respectively. The former reported a complex maser spot distribution with a maximum extent of 230 mas (= 54 au) in the east-west direction. The latter confirmed the overall spot distribution and determined an average shell diameter of $\sim$164 mas ($\sim$38 au). 

VLBI observations of the \water\ masers of R~Crt during our monitoring programme were made by \cite{imai97a} in 1994 and 1995, \cite{migenes99} in 1996, and \cite{ishitsuka01} in 1998. While previous VLA observations (\citealt{bowers94}; \citealt{colomer00}) were not able to resolve the individual maser spots,  Imai et al. reported compact maser spots with sizes $<1$ mas ($<0.2$ au). \cite{ishitsuka01} could trace proper motions for several of these compact maser spots. The movements indicate the presence of a bipolar outflow. According to their model, the \water\ maser shell extends in radius between 12 and 24 au.
Between these radii the outflow velocity increases from 4.3 to 7.4 \kms. As the outflow velocity at the outer border is only $\sim$75\% of the highest outflow velocity observed by us ($\sim$10 \kms), the outer parts of the \water\ maser shell were not detected by \cite{ishitsuka01}, either because of the weakness of the maser emission there or because the emission was resolved out. 
To predict the distance at which the stellar wind reaches $\sim$10 \kms, we used the velocity gradient found by \cite{ishitsuka01} for the inner parts of the shell. The extrapolation gives an outer radius of the \water\ maser shell of $\sim$35 au, about 30\% larger than observed with the VLA in 1988 \citep{bowers94}. 
Recent VLBI measurements carried out in May 2015 and January 2016 (\citealt{kim18}) found an asymmetric one-sided outflow structure based on the location of all \water\ maser emission sites in a $100 \times 80$ mas$^2$ ($20 \times 24$ au$^2$) region south of the stellar position. This observation puts the assumption of a spherical outflow of the \water\ maser shell of R~Crt into question and this is discussed in Sect. \ref{cse-model}.

 The observations of \cite{colomer00} took place during the rise of the first burst which we detected, peaking in October 1990. They found the spatial component belonging to this burst close to the line of sight to the adopted stellar position, which indicates that the burst feature originated from a spatial component moving radially away from us in the rear part of the \water\ maser shell. The later bursts we discussed, those detected in 1992 and 2002--2004, were not covered by interferometric observations. 

Unlike for bursts, the association of spectral maser features with their spatial counterparts in single maps is often not possible because of the presence of several spatial candidates with similar velocities. In addition, due to the short timescales on which individual maser components survive in the \water\ maser shell of R~Crt, also a reliable spatial association of maser components between maps observed several years apart, is difficult. Unlike for the longer living maser components in the much larger shells of RSGs  (Brand et al., in preparation), the use of maser features from our single-dish spectra to connect spatial components in the contemporaneous maps of R~Crt taken between 1988 and 1998 was unfeasible for the same reason. 

\begin{figure*}
\resizebox{18cm}{!}{
\includegraphics{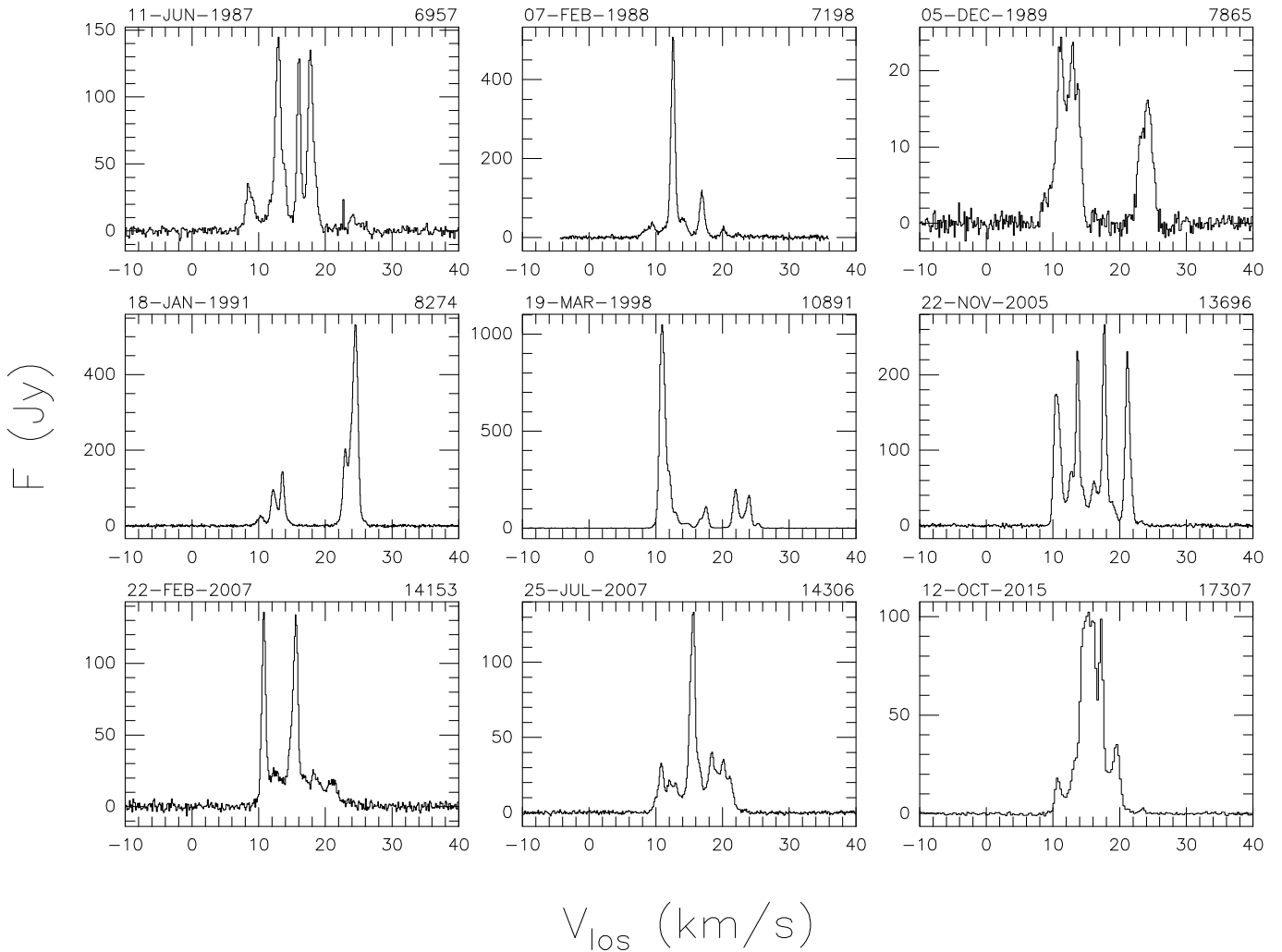}}
\caption{Selected H$_2$O maser spectra of RT~Vir. The calendar date of the observation is indicated on the top left above each panel, the TJD (JD-2440000.5) on the top right.
}
\label{fig:rtvir_sel}
\end{figure*}

\subsection{Summary}
The variability pattern of the \water\ masers of R~Crt is dominated by short time fluctuations on the scale of several months. Frequently, the brightness in individual velocity intervals increases to such a level (bursts) that they dominate the integrated maser emission output of the star. The variations are not entirely random. There are year-long phases where the overall emission level changes and there are velocity intervals where the general emission level is systematically higher than in others. The latter result indicates that the \water\ maser shell of R~Crt is not homogeneous, but contains regions in which \water\ maser emission is preferentially excited. These regions persist over decades. 

The velocity range over which maser emission is detected in individual spectra varies depending on the brightness of the emission at the most blueshifted and redshifted velocities.
In a standard CSE, emission at these velocities comes from the outer periphery of the \water\ maser shell and in these regions, the maser may not always be excited or strong enough to be detected. Overall, the outflow velocity within the \water\ maser shell reaches $V_{out}^{max}=10$ \kms, which is $\sim$85\% of the final expansion velocity $V_{\rm exp}$. 

The interferometric observations showed that the strongest maser emission is coming from a distance of 12--24~au from the star (\citealt{colomer00}; \citealt{ishitsuka01}), but we argued earlier (Sect.~\ref{since87}) that weaker maser emission may be present outside this region with distances as far as $\sim$35~au. 
The crossing time for the  \water\ maser shell of R~Crt, adopting a maximum shell thickness of $\sim$23~au and an average outflow velocity of $\ga$6~\kms,\ would take $\la$18 years, so that over the duration of our monitoring programme, a complete exchange of the matter in the \water\ maser shell took place.

\section{RT~Vir \label{rtvir-main}}
RT~Vir is a semi-regular variable AGB star with a distance of $226\pm7$ pc (\citealt{zhang17}; see also \citealt{imai03}), determined from a trigonometric parallax measurement of the \water\ masers.
We adopt as radial velocity of the star $V_{\ast} = 17.3\pm0.2$ \kms\ and as final expansion velocity $V_{\rm exp} = 9\pm1$ \kms, as determined from circumstellar CO and 1612 MHz OH emission (see Table~\ref{centralcoords}).

\water\ maser emission in RT~Vir was first detected in 1973 by \cite{dickinson76} with a single feature of moderate brightness at 15.1 \kms\ (these observations had an upper limit for non-detections of 5-6~Jy). \cite{cox79} monitored the maser for 1.5 years in 1975-1977 and detected a much brighter feature ($\approx$200 Jy) at 22 \kms. VLBI observations in 1976 by \cite{spencer79} determined the size of this feature at the level of 1.6 mas ($\approx$0.4 au). Both known features were monitored by \cite{berulis83} between 1980 and 1982. They found stronger variations for the 22 \kms\ feature compared to those at 15 \kms. Additional emission between 12 and 15 \kms\ was detected by \cite{spencer79} and \cite{berulis83}. Over the course of 1.5 years (1984-1986). \cite{berulis87} detected seven bursts of maser features in a broad velocity interval ($11.4 < V_{\rm los} < 24.3$ \kms) with a maximum flux density of 2400 Jy at 12.7 \kms\ in March 1985. This feature was dominating the spectra also at the beginning of our monitoring programme in 1987. Thus, between its discovery in 1973 and the year 1987, the dominating maser features appeared at $\sim$15, $\sim$22, and later at $\sim$13 \kms, and persisted for many years.

New interferometric observations were obtained in 1-2/1985 by \cite{bowers93} and \cite{yates94}, and in 12/1988 by \cite{bowers94}, while the  $\sim$13 \kms\ feature dominated the emission. Maser emission was seen over a velocity interval 6.5-26.4 \kms. Emission, albeit weak, at $V_{\rm los} < 10$ and $V_{\rm los} > 24$ \kms\ was detected for the first time. \cite{bowers93} found an elongated distribution of the maser spots in the northwest-southeast direction and a shell radius of $\approx$70 mas ($\approx$16 au). The distribution of the spots changed between the beginning of 1985 and the end of 1988, but the overall extent remained the same  (\citealt{bowers94}). \cite{yates94} favoured a geometrically thick shell with inner and outer radii of 50 and 62 mas, respectively.

\begin{figure}
\resizebox{9cm}{!}{
\includegraphics
{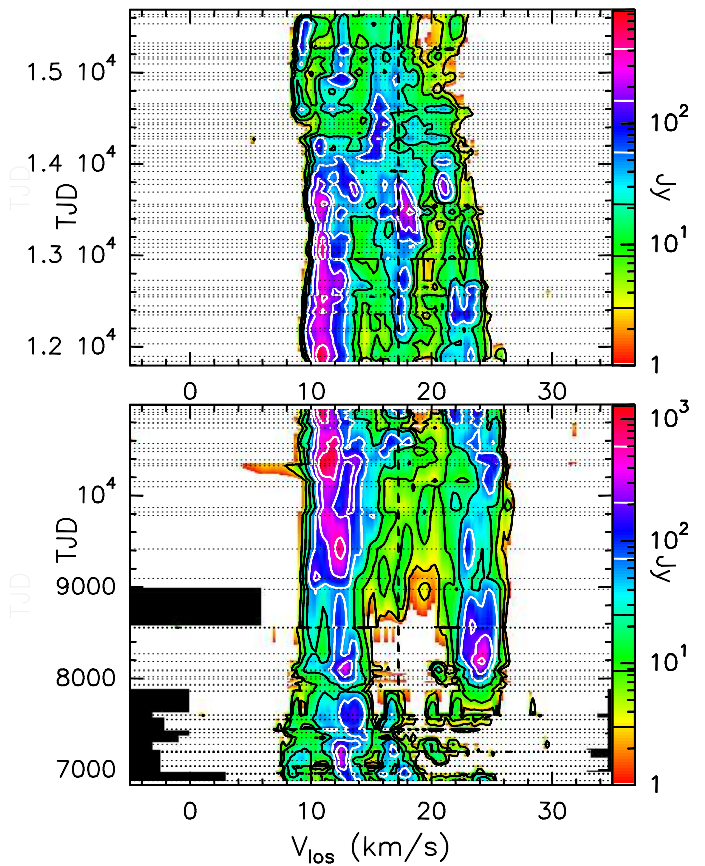}}
\caption{As in Fig.~\ref{fig:rcrt-fvt}, but for RT~Vir. Lower panel: First spectrum: 1/4/1987; JD = 2446886.5, TJD = 6886 and last spectrum 11/5/1998. Upper panel: First spectrum 27/10/2000 and last spectrum 20/3/2011. The black areas in the plot indicate the unobserved parts of the spectrum on those particular days. The low-level emission at V$_{\rm los} \sim$4-10~\kms\ at TJD=10329 is an artefact. 
}
\label{fig:rtvir-fvt}
\end{figure}

\begin{figure}
\resizebox{9cm}{!}{\rotatebox{270}{
\includegraphics
{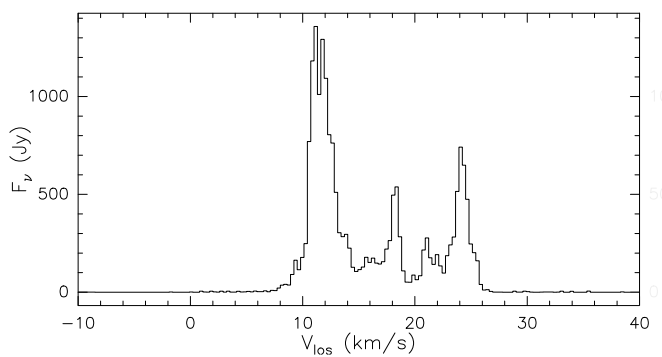}}}
\caption{Upper envelope spectrum for RT~Vir; 1987-2015 (cf. Fig.~\ref{fig:rcrt-upenv}).}
\label{fig:rtvir-upenv}
\end{figure}

\begin{figure}
\resizebox{9cm}{!}{\rotatebox{270}{
\includegraphics
{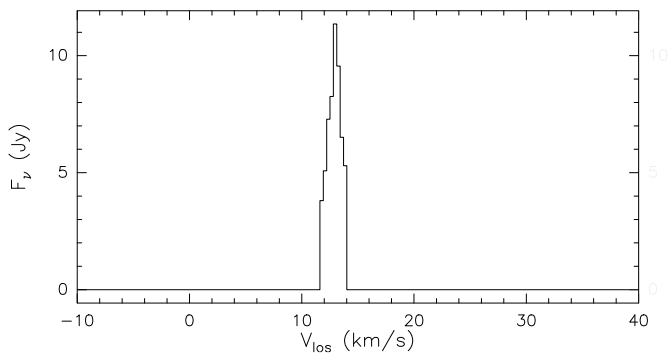}}}
\caption{Lower envelope spectrum for RT~Vir; 1987-2015 (cf. Fig.~\ref{fig:rcrt-loenv}).}
\label{fig:rtvir-loenv}
\end{figure}

\begin{figure}
\resizebox{9cm}{!}{\rotatebox{270}{
\includegraphics
{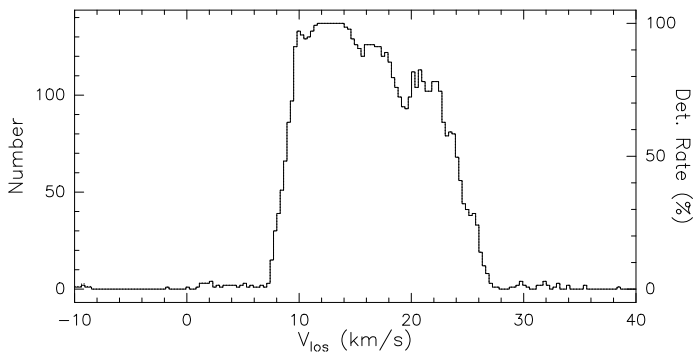}}}
\caption{Detection-rate histogram for RT~Vir; 1987-2015 (cf. Fig.~\ref{fig:rcrt-histo}).}
\label{fig:rtvir-histo}
\end{figure}

\begin{figure}
\resizebox{9cm}{!}{\rotatebox{0}{
\includegraphics
{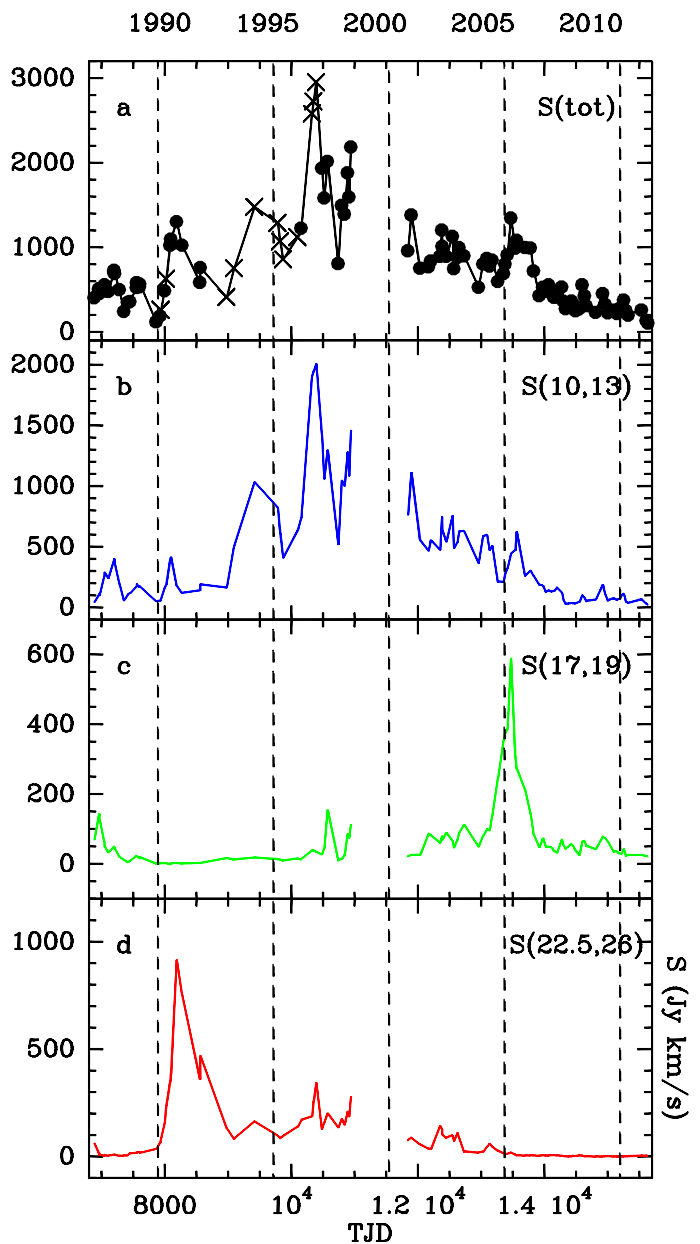}}}
\caption{\water\ maser light curves of RT~Vir in 1987-2011, showing the integrated flux density $S$ in different velocity intervals versus TJD (= Julian Date - 2440000.5). Data points on either side of the gap in the observations (May 1998 - Nov. 2000) are not connected. Spectra taken within three or four days  were averaged before $S$ was determined. The vertical lines connect the first day of each of the years indicated at the top of the figure with the  corresponding TJD on the lower axis. {\bf a.} $S(tot)$ is the total flux of all maser emission, between $6.4 < V_{\rm los} < 28$ \kms. Observations marked with crosses are from Effelsberg, the dots mark Medicina data. {\bf b.--d.} $S(10,13)$, $S(17,19)$, $S(22.5,26)$ are the  integrated fluxes densities in the velocity intervals indicated.
}
\label{fig:rtvir-lcurve}
\end{figure}

\subsection{Results from the  \water\ maser monitoring programme  \label{rtvir-campaign}}

\subsubsection{Profile characteristics and velocity ranges}
Representative maser spectra of RT~Vir from our observations between 1987 and 2015 are shown in Fig.~\ref{fig:rtvir_sel}, while all spectra are collected in Fig.~\ref{fig:rtvir_all} (Appendix). In between (5/1998--10/2000), the observations were interrupted for $\sim$2.5 years. Therefore, the FVt-plot (Fig.~\ref{fig:rtvir-fvt}), giving an overview of the behaviour of the maser emission, is split into two parts. The emission pattern is similar to the one observed for R~Crt, with brightness variations on several timescales. In addition to fluctuations lasting months to more than a year, a much longer-living feature is also present, at 11 \kms, which was strong for many years and which dominates the blue part of the FVt-plot. 

The upper envelope spectrum (Fig.~\ref{fig:rtvir-upenv}) and the detection-rate histogram (Fig.~\ref{fig:rtvir-histo}) show emission in the range $7.4 < V_{\rm los} < 27.0$~\kms\ ($\Delta V_{\rm los} = 19.6$ \kms), which encompasses all the maser feature velocities seen ever, except for weak emission extending down to $V_{\rm los} =6.5$ \kms\ detected by \cite{bowers94} in 1988. The maximum outflow velocity in the \water\ maser shell is found to be $V_{\rm out}^{max} = 9.0\pm0.5$ \kms.
\water\ maser emission is therefore emitted over the complete velocity range $V_{\ast} \pm V_{\rm exp}$, indicating that for RT~Vir the outflow velocity reaches the terminal expansion velocity already in the \water\ maser shell. As in R~Crt, emission was not detectable at all velocities of the \water\ maser velocity range all the time. The maximum velocity range is centred on the stellar velocity ($\mid V_{\rm c}-V_{\ast}\mid = 0.1$ \kms), which is consistent with a spherically symmetric  \water\ maser shell.

The three peaks in the  upper envelope spectrum are the long-lasting feature at 11~\kms\ and the emission features at 18 and at 24 \kms, which became strong only for a short duration (see below). The lower envelope spectrum (Fig.~\ref{fig:rtvir-loenv}) shows that only the 13~\kms\ feature was present permanently during the monitoring programme. However, the 13~\kms\ feature was dominating the emission profile only in the first years until 1990. 

As has been found for R~Crt, also RT~Vir shows velocity range variations in the FVt-plot (Fig. \ref{fig:rtvir-fvt}). Before 1989 (TJD $<$7500) there was little, if any, emission detected at $V_{\rm los} > 18$~\kms\ and in the period 1/1990 --12/1991 (TJD $\approx$ 7900 -- 8600) no or little emission was detected between 15 and 22~\kms, giving the profiles a double-peaked shape (Fig. \ref{fig:rtvir_sel}). From a (very low-noise) Effelsberg spectrum taken in May 1990 (TJD = 8023), we find, however, that there is actually emission throughout this velocity interval with a minimum of $\approx$1~Jy between 17 and 20~\kms, but well below the sensitivity of the FVt-diagram. The red border of the velocity range shows an eye-catching curvature, which is due to a continuous fading of the emission at velocities $V_{\rm los}>22$~\kms. The asymmetry of the emission over the velocity range is reflected also in the detection-rate histogram (Fig.~\ref{fig:rtvir-histo}) showing that emission in general was less frequently detected in the red part ($V_{\rm los} > 15$~\kms) of the maser profile compared to the blue part. As in R~Crt, the velocity range variation is due to a temporary weakness of the maser emission at the most redshifted velocities.

\subsubsection{The \water\ maser light curve 1987-2011}
Figure~\ref{fig:rtvir-lcurve} shows in the upper panel the radio light curve of RT~Vir  of the \water\ maser emission $S(tot)$ as obtained by determining the integrated flux in the velocity interval $6 < V_{\rm los} < 28$~\kms. 
At the beginning, the light curve showed a steady increase and the integrated flux rose until $\sim$1997 (TJD$\sim$10500) from $S(tot)\approx550$ to  $S(tot)\approx3000$ Jy~\kms . After our observing gap, which ended in 2000, the flux continuously decreased down to $\approx110$ Jy~\kms\ in 2011. In 2015 (not shown), the flux had recovered to $S(tot) = 440$ Jy~\kms\ on average. Very weak phases with peak flux densities $<30$ Jy ($S(tot) < 200$ Jy~\kms) were recorded between December 1989 and February 1990 (TJD$\sim$7900) and between February and March 2011 (TJD$\sim$15600). just before the end of the monitoring programme. On top of the long-term trend, strong brightenings of maser features on timescales $\sim$1 year were observed frequently and over all parts of the velocity range. Due to this behaviour, the maser profile varied strongly over time (cf. Fig. \ref{fig:rtvir_sel}). 

The other panels of Fig. \ref{fig:rtvir-lcurve} show light curves restricted to the integrated flux density, $S$, in velocity intervals $V_{\rm los}$=10-13, 17-19, and 22.5-26~\kms,\ encompassing several maser features, which dominated the overall emission for years, as is evident in the FVt-plot (Fig.~\ref{fig:rtvir-fvt}). The outstanding characteristic of the maser profile during our monitoring programme was the feature at 11~\kms, which dominated for about 12 years between 1993 and September 2004 (TJD $\sim$9000 -- $\sim$13200) with flux densities of several hundred to more than a  thousand Jy. Flux densities reached a maximum of $\sim$1400~Jy between September and November 1996 (TJD $\sim$10300) and steadily declined afterwards, with some fluctuations. The red wing of the feature is blended with weaker maser features at velocities up to $\sim$14~\kms. Together, these features create the prominent bright ridge in the blue part of the emission pattern displayed in the FVt-plot. 

The 11~\kms\ maser feature in itself is a blend as judged from the velocity centres determined by a line fit to the feature varying between 10.6 and 11.9~\kms, from the asymmetry of the line profile, and from the occasional presence of two peaks within the feature. Between January 1998 and July 2005, a single line prevailed with the velocity centre  permanently at $11.0\pm0.2$~\kms. This is the velocity at which  the 1996 burst also occurred. The 11.0~\kms\ line is distinguishable in the profiles until July 2007 (TJD$\sim$14300).

After 2004 and until the end of 2005, a strong feature at $\sim$18~\kms\ was present, which dominated the profile for a couple of months (Fig.~\ref{fig:rtvir-lcurve}, panel c). In the years thereafter, the strongest emission feature was seen at $\sim$15.5~\kms\ (2007/2008; TJD = 14250 -- 14700). Also at the edges of the maser velocity range features could dominate the profile for some time. This was the case in 1990/1991 at $\sim$24~\kms\ (TJD = 8100 -- 8600) (Fig.~\ref{fig:rtvir-lcurve}, panel d) and in 2009/2010 at $\sim$9~\kms\ (TJD = 15050 -- 15550), while all the other features were relatively weak. Except for the 11 \kms\ feature, the  duration for individual features to dominate in the spectral profile is $\la$1.5 years. 

We analysed the optical and maser light curve of RT~Vir for periodicity and correlations between them. The V-band optical light curve between 1987 and 2015, which shows small amplitude variations, was obtained from AAVSO. A periodogram analysis revealed a peak at $\sim$370 days, and not at the commonly adopted period of 155 days (\citealt{samus17}). 
A 370-day optical period was also found by \cite{imai97b}, \cite{shintani08} and \cite{zhang17}. 
We should note, however, that the phase coverage of the optical light curve is rather incomplete because of the coincidence of this period with Earth's orbital period. 
Due to the small amplitude variations of the optical light curve and the irregular fluctuations of the maser emission, no correlation between them was found. A periodogram analysis revealed also no periodicity in the maser light curve.

\subsection{Comparison with the single-dish Pushchino and VERA-Iriki monitoring programmes}
Our monitoring programme overlapped in part with the Pushchino monitoring programme (\citealt{berulis83, berulis87}; \citealt{mendozatorres97}; \citealt{lekht99}; hereafter Pushchino group) carried out between 1980 and 1998. Combining both programmes, the \water\ maser of RT~Vir was followed regularly, with just a few gaps, for about 35 years. 
The Pushchino group already found \water\ maser emission over exactly the same velocity range as we had surveyed, along with average profiles similar to our upper envelope spectrum (Fig.~\ref{fig:rtvir-upenv}) with three peaks in the velocity intervals 11--14, 16--18, and 22--25 \kms. They noticed the dominating role of the $\sim$24 \kms\ feature in 1990/1991 and the three-month-long weak phase in 1989-1990. As in our light curve (Fig. \ref{fig:rtvir-lcurve}), they found an overall maximum in 1996--1998 (TJD $\sim$10000 -- 11000). They associated the dominating maser features with bursts having a mean duration of 0.5 years. The exception is the 11 \kms\ maser feature, which appeared in 1996 in their observations and persisted until the end of their observations.

Interestingly \cite{lekht99} measured on 1/6/1998 an exceptionally strong peak flux of $\sim$2400 Jy for the $\sim$11 \kms\ feature. This burst was missed by us because our monitoring programme was interrupted after May 1998 for about 1.5~years. But also our observations (cf. Fig.~\ref{fig:rtvir_all} in the Appendix) confirm the strong flux increase starting beginning of 1998. The decline of this burst, unfortunately, was not covered by either monitoring programme.

Using Gaussian fits to analyse the individual maser features, velocity shifts of up to 1 \kms\ were found by the Pushchino group, which could be followed over typically $1.0\pm0.5$ years. Few features were found with lifetimes of two and more years (\citealt{lekht99}). The group note that the features are not necessarily different spatial components, but may reflect activity periods of longer-living components. Some peaks are blends of maser features having similar radial velocities. 

The VERA-Iriki monitoring programme (\citealt{shintani08}) observed RT~Vir's \water\ maser biweekly between May 2003 and September 2005 ($\sim$12800 $<$ TJD $< \sim$13600). They reported periodic variations for individual maser features with a period $P\sim$300 days (thus somewhat similar to the $\sim$370 days found from the optical data). 
A fit with a sine-function to the integrated flux density using the GCVS period of 155 days resulted however only in a bad solution. 
The inconclusive results on periodicity obtained from their 2.5 years of  observations do not, thus, stand in contradiction to our failure in finding periodicity in the \water\ maser variations over a much longer time range.
One-time single-dish observations by \cite{takaba01} in 1991 and \cite{kim10} in 2009 of the \water\ masers during our monitoring programme showed velocities and profiles  that are consistent with our observations.

\subsection{Interferometric observations since 1987 \label{rtvir-interferometry}}
Interferometric observations during our monitoring programme were carried out in 1994 and 1996 with MERLIN (\citealt{bains03}; \citealt{richards12}); in $1994-1996$ using VLBI (\citealt{imai97a, imai97b}), and in 1998, 2009, and 2014, using the VLBA (\citealt{imai03}; \citealt{lealferreira13}; \citealt{zhang17}, respectively). In 1994, Bains et al. found the strongest emission in a ring with radius of 75 mas ($\sim$17\,au). While the overall size of the maser region was similar to those determined previously, matches with individual spatial components were not possible. In addition to the emission in the ring, faint maser components were detected north and south of the ring with a separation $\le 370$ mas. This indicates that the outer border of the \water\ maser shell may extend at least up to 40 au. 

The VLBI observations of 1994/1995 by \cite{imai97a} detected compact maser components with a size $\le1.3$ mas, coinciding in velocity with the strongest maser features seen in contemporaneous single-dish spectra. The components were distributed in an east-west direction over a range of 160 mas, which is compatible with the size of the ring deduced by \cite{bains03} from their 1994 map. The compact components were monitored by \cite{imai97b} over five months in 1996 and significant line-of-sight velocity shifts could be detected. 
Contrary to the velocity shifts of spectral features seen in single-dish spectra, the VLBI-determined shifts were not affected by blending of more than one spatial component with similar velocities, as many maser components are resolved out by VLBI and the remaining spatial components are well separated in velocity. However, the shifts could not be explained by simple kinematic models. Later, with the improved resolution provided by the VLBA, \cite{imai03} were able to show that radial velocities and proper motions of their 1998 observations were indeed compatible with a  CSE that is expanding roughly spherically.

Because of the frequent interferometric observations RT~ Vir's \water\ maser emission it became evident that shell boundaries cannot be considered as constant. Between observations separated by many years they vary by factors of $\sim$2 \citep{lealferreira13}. In addition to effects due to instrumental resolution, this is likely due to the transient nature of the masers, which illuminate the underlying emission region often only in part. An example is the frequent lack of or often weak emission at the most redshifted velocities of the \water\ maser velocity range discussed above (Sect. \ref{rtvir-campaign}). 

In summary, the \water\ maser shell can be described as having an inner boundary of 4--10 au (\citealt{bains03}; \citealt{imai03}; \citealt{richards12}; \citealt{lealferreira13}), a major ring-like region between 10--20~au (\citealt{imai97b}; \citealt{bains03}; \citealt{richards12}), where the strongest maser emission comes from, and an outer region with occasional detections, which can extend out to 45~au (\citealt{bains03}; \citealt{imai03}). The simple expanding shell model is however oversimplified, as is evident from the detailed spatial distributions of the masers given that the emission regions at the most extreme velocities projected on the sky do not coincide and an east-west velocity gradient was observed for some time (\citealt{bowers93}; \citealt{bains03}; \citealt{richards12}).

\subsection{The long-living 11 \kms\ maser feature \label{sec:11kms}}
The prevailing maser feature in single-dish spectra since April 1993 at $\sim$11 \kms\ was first mapped in 1994 by \cite{bains03} and was seen to be located north-west from the adopted stellar position. It was found in the same direction and at similar distances, and was dominating also in the six MERLIN maps taken in 1996 by \cite{richards12} and the five VLBA maps taken in 1998 by \cite{imai03}. Although no further interferometric observation is available until the feature became indistinguishable in the line profiles in 2007, we conclude that this maser feature was always emitted by the same sector in RT Vir's shell.

As in the case of R~Crt, individual maser components have rather short lifetimes and often inhibit a reliable association of spatial maser components between maps observed several years apart. The longevity of the 11 \kms\ feature implied by its $>$14 years (1993 -- 2007) presence in the profiles is, therefore, surprising. Before 1998 and after 2005, the feature was a blend of several lines, while between 1998 and 2005 ($\sim$7.5 years,) the feature had a constant velocity (variations $<0.06$ \kms~yr$^{-1}$), with no evidence in the profile for more than one maser line making a contribution. In principle, the 11 \kms\ feature could originate in a region where several shorter-lived maser components with similar velocities mimic a long-living maser component. However, we found no velocity variations that could be used as an indication of multiple clouds replacing each other over time. Therefore, we conclude that the feature was emitted by a single cloud with an exceptionally long lifetime.  

If the \water\ maser shell is a zone in the stellar wind where matter is still being accelerated, then the absence of any velocity shifts over many years is hard to explain in the case of a single, long-living cloud. The maser cloud could move perpendicularly to the line of sight and would then have a velocity coincident with the stellar velocity ($V_{\rm los} = V_{\ast}$). This is not the case for the 11 \kms\ feature. The other possibility would be a stationary region in the shell, where \water\ molecules would be excited while passing through it, mimicking a long-lived maser cloud. Considering the fluctuation time scales of maser features of the order of months and the frequently observed velocity shifts of \water\ maser components in the shell of RT~Vir, it is hard to believe that such a region with stationary maser excitation conditions can exist on timescales an order of magnitude longer than observed generally for individual maser clouds in RT~Vir's shell. 

We conclude, therefore, that the 11 \kms\ feature is emitted by a maser cloud moving in a part of the shell, which must have a constant outflow velocity. This part is probably the outer part of the shell, as we found that in RT~Vir the terminal expansion velocity is already reached within the \water\ maser shell (Sect. \ref{rtvir-campaign}). The absence of a velocity gradient in at least part of the shell seems to contradict \cite{richards12} finding a rather steep velocity gradient, $K_{\rm grad} \approx 0.4$~\kms~au$^{-1},$ in the \water\ maser shell of RT~Vir.
In 1996, the 11 \kms\ maser component had an angular separation from the adopted position of the star of $\sim$32~mas \citep{richards12} and of $\sim$46~mas in 1998 \citep{imai03}, that is, the projected distance between this component and the star was 7.2 and 10.4~au, respectively. The 11 \kms\ feature has a projected outflow velocity (w.r.t. $V_{\ast}$) of 6.3 \kms; it is in the front part of the shell approaching the observer. We concluded that because of its constant velocity, it moves at $V_{\rm exp} = 9$~\kms, and, thus, the outflow direction makes an angle of $\sim$46\degs\ with the line of sight. 
Furthermore, the  projected distance of 10.4~au in 1998 corresponds to a real distance of 14.6~au. 

A location of the 11 \kms\ cloud at $r\sim$15~au in 1998 and its continuous movement since then with constant velocity implies that the acceleration in the inner shell is already completed  at that distance from the star. Until 2007 (the year, up to which the 11 \kms\ feature was distinguishable), the cloud would have moved $\sim$18~au, arriving at a distance from the star of about 33~au, which agrees with the outer radius found by \cite{richards12} for 1996 when scaled to the distance used here. 
The general decrease of the flux density between 1996 and 2007 may then be a result of the approach of the 11 \kms\ cloud to the outer boundary of the shell, where the excitation conditions deteriorate. 
On the other hand, assuming an outer radius at 42--45~au (\citealt{bains03} and \citealt{imai03}, respectively, for 1994 and 1998) and reasoning backwards, we find that if the 11~\kms\ component was at those distances in 2007, it would have been at 24--27~au from the star in 1998. In summary, the wind-acceleration zone is thus restricted to a distance of 15--27~au from the star.

Perhaps more remarkable than a long-living cloud in the water maser zone of the CSE, which has already been considered a likely possibility by \cite{bains03} and \cite{richards12}, is the existence of a long-living maser originating from that cloud. The masing region had managed to maintain the appropriate conditions for maser excitation because the cloud moved outside the acceleration zone, making it possible, for example, to conserve velocity coherence.

This explanation for the absence of a velocity shift of the 11 \kms\ feature implies the possible presence of more maser clouds with constant velocity moving elsewhere in the outer part of the \water\ maser shell. We have not found other convincing cases, possibly because the relative brightness of the 11 \kms\ feature was outstanding, allowing a precise determination of its velocity over a long period of time. The longevity may imply that the lifetimes of the maser components drastically increase, as soon as they leave the acceleration zone in the CSE. 

\subsection{Summary}
The \water\ maser variability pattern of RT~Vir is dominated by short-lived fluctuations, superimposed on decade-long changes of the average maser brightness. The fluctuations include bursts in particular velocity intervals, with lifetimes of the order of several months, as has been found also in R~Crt (Section \ref{rcrt-main}) and reported for RT~Vir by \cite{lekht99} before. As in R~Crt, the variations in total maser brightness are asynchronous superpositions of emission features coming from different velocity intervals, where individual features are able to dominate the maser brightness on timescales of $\la1.5$ years.
Expected velocity shifts of individual features were strongly affected by blending also in RT~Vir. It is only for the 11~\kms\ component that the (absence of a) velocity shift could be studied over about half of the time when it could be distinguished as a feature. 

The \water\ maser brightness of RT~Vir 1987--2015 was characterised by one bright emission region responsible for the  11 \kms\ feature in the north-western part of the shell. No systematic shift in velocity was observed, indicating that the feature moved with an almost constant velocity, that is, the velocity gradient in the shell must have been practically absent. No random  variations in velocity were observed either, which could be interpreted as being due to multiple short-lived clouds making up a long-lived region. Instead, we concluded that the 11 \kms\ feature originated from a single cloud traversing about half of the shell, moving in its outer part.

The dimensions of the \water\ maser shell of RT~Vir are comparable to those of R~Crt and also RX~Boo (Paper I) with inner and outer boundaries of $\sim$5~au and 45~au, respectively. Adopting a shell thickness of  $\sim$40 au and an average outflow velocity of $\sim$8 \kms, the crossing time of the shell is $\sim$24 years.

Also, RT~Vir shows variations of the \water\ maser velocity range over time and, as in R~Crt, this is a consequence of the brightness variations of the emission at the most blueshifted and redshifted velocities. According to the maximum velocity range, the final expansion velocity V$_{\rm exp}$ of RT~Vir's wind is already reached within the \water\ maser shell.

\section{\water\ maser properties in semi-regular variable stars \label{discussion}}
\subsection{Short-term variability and long-living structures}
We found compelling evidence that the \water\ maser brightness variations in R~Crt and RT~Vir are a superposition of two types of variations with different timescales. There are short-term fluctuations on timescales $\la$1.5 yr, and long-term variations on timescales of decades. A third type, the response of the maser brightness to (periodic) luminosity variations of the stars is not seen, apparently because of the absence of periodic variations (R~Crt) or relatively small amplitudes (RT~Vir).

The short-term variability pattern of the \water\ maser emission of R~Crt and RT Vir is consistent with the pattern we found for the SRVs RX~Boo and SV Peg (Paper I). As found before by the first monitoring programmes (\citealt{cox79}; \citealt{berulis87};
\citealt{lekht99}), the brightness of maser spectral features fluctuates, increasing occasionally by large factors and developing into bursts. Such occasional  increases in brightness might be caused by a sudden increase in the coherent path length in the masing cloud (e.g. \citealt{reid90}), an increase in the pumping  efficiency or a change in beaming angle, for example. Bursts can also be caused occasionally by amplification of background emission, as has been convincingly shown by \cite{richards12}. The lifetime of spectral features is difficult to estimate as features often blend with each other in velocity space. Bursts surpass flux density levels of other spectral features for a couple of months up to more than a year. This gives a typical timescale of several months for the lifetime of spectral features, which is considered as the time during which a spectral feature can be distinguished in the spectral profiles. These lifetimes are consistent with those reported before for SRV and Mira variables (e.g. \citealt{lekht99}; \citealt{shintani08}; \citealt{richards12}). The shorter lifetimes compare well also with the time ranges of detectability ($\la6$ months) of spatially resolved maser spots observed by \cite{imai97a} and \cite{imai03}. In fact, there is a range of lifeftimes, but only few were found surpassing two years (RT~Vir: \cite{lekht99}; RX~Boo: Paper~I). \cite{bains03} connect them to the sound-crossing times of individual clouds. In the case of SRVs (and Mira variables) the sizes are 2--4 au, from which they infer upper limits to the lifetimes of $\la2$ years. 
Sometimes masers can persist for a very long time, as demonstrated by the case of the 11~\kms\ feature in RT~Vir (Sect.~\ref{sec:11kms}). The clouds in which the masers originate live at least as long as the masers themselves, therefore, the cloud lifetimes may also be much longer than a few years. Whether this is true for cloud lifetimes in general, as suggested by \cite{richards12}, who argued for lifetimes of the order of the water maser shell-crossing time, remains to be seen. To get an insight into this possibility, it would be necessary to carry out a VLBI experiment over a much longer timeframe than currently available. This would allow us to see whether maser spots switch on and off, leading to a systematic shift in velocity and following a trajectory that is consistent with a radially outflowing cloud.

The maser emitting clouds are located in a shell of favourable excitation conditions (the \water\ maser shell) within the CSE with an inner radius of a few au and an outer radius a factor of ten larger. In general, the \water\ maser shell is located at radii where the stellar wind is still accelerated. The inner boundary is probably due to the collisional de-excitation of the maser attributed to high densities inside the boundary (e.g., \citealt{cooke85}; \citealt{reid90}) and the outer boundary is probably due to dissociation of the \water\ molecules  (\citealt{goldreich76}). The clouds are considered to be 'density bounded', meaning that they have higher densities than their surroundings. If they were to remain undestroyed they would traverse the shell on timescales of two to three decades and, in general, would increase their velocity by a factor of two to three (\citealt{bains03}; \citealt{richards12}). However, the clouds are emitting the masers into the line of sight only for a couple of months \citep{imai03} and there are several reasons for these short lifetimes. The clouds could change their beaming direction or simply switch on and off because of velocity fluctuations in the stellar wind, which induce fast changes of the velocity coherence length towards the observer. Alternatively, the clouds may form and dissolve on such timescales due to density fluctuations in the wind. Under this hypothesis, individual emitting maser clouds trace only a small part of the trajectory of the stellar wind through the \water\ maser shell.

For an isotropic stellar wind traversing the \water\ maser shell, the maser clouds are expected to appear and disappear in random directions, and a random emission pattern without preference of particular velocity intervals is expected to be seen in the FVt-plots. Instead, we are observing 'long-living structures', that is, a preferential excitation of masers in particular velocity intervals over timescales of ten years and more. We propose that these structures come from regions in the shell that have sizes much larger than individual clouds. They may have higher densities than the average density in the shell, making the probability of the formation of individual maser clouds higher than in other parts of the shell. This would explain the observed repeated brightenings of maser components at similar velocities. These regions responsible for the 'long-living' structures may be extensions of density enhancements in the stellar wind close to the stellar surface due the departure of large convection cells from the photosphere (see Section \ref{cse-model}). Because of their size, they will have longer lifetimes than individual clouds and might be able to cross the shell undestroyed. They would then be detectable with timescales of decades corresponding to the shell-crossing times in SRVs, until they eventually move out from the \water\ maser shell. Longer living structures were not addressed in Paper~I, but a long lasting decline of the light curve of RX~Boo over $\sim$7 years (1987--1993) indicates that their presence cannot be excluded there either. Such structures were anticipated by \cite{richards12}, who proposed the existence of 'long-living clouds' to explain the reappearance of masers at similar velocities and positions after a couple of years.

A special case is the outstanding 11 \kms\ feature in RT~Vir with a lifetime $>$14 years. It was observed to move at a constant velocity for $>$7.5 years indicating that the cloud or region had already left the acceleration zone of the CSE. Because of the absence of a velocity gradient, we cannot exclude that the 11 \kms\ feature was emitted by multiple clouds within a region, with all of them having the same velocity. However, the extremely constant velocity indicates that the 11 \kms\ feature originated more likely from an individual and unusually stable maser cloud. Maser clouds with lifetimes of many years have been observed, for example, in RSGs (\citealt{murakawa03}; Brand et al., in preparation), where the \water\ maser shells extend to much larger radii than in SRVs or Mira variables. A decrease, and ultimately absence, of acceleration of the stellar wind in the outer parts of the \water\ maser shells may inhibit the perturbation of individual clouds that would otherwise lead to the suppression of the maser within months.

A similar explanation related to mass-loss variations may hold for the cause of the long-term variations of the maser brightness, with phases of brightenings and dimmings, seen in R~Crt and RT~Vir on timescales of several years to decades. With typical shell-crossing times of $\la25$ years, it is tempting to associate these long-term variations with changes of the stellar wind properties within the shell, possibly connected to global variations of the mass-loss rate or temporary directional increases of the mass-loss due to the convection processes in the stellar photosphere (see below in Section \ref{cse-model}). Variations of the average brightness level would then occur on timescales of the shell-crossing times.
For example, the decade-long decrease in average brightness observed for R~Crt from the start of the monitoring programme until 2002 could be connected, for example, to the transit of matter in the shell with reduced density of water molecules caused by a decrease of the mass-loss rate. Similarly, the recovery of the average brightness afterwards could have been the start of a new episode of higher mass-loss rates. Because of the travel times between stellar photosphere and the inner boundary of the water maser shell, these variations in mass-loss rate must have happened more than a decade before the effects would show up in our maser observations.

\begin{table*}
\caption{\water\ maser luminosities, stellar luminosities, and mass-loss rates of the sample. The stellar type is semi-regular variables (SRV).
Column "D" lists distances with their references as noted below the Table. 
Characteristic levels of \water\ maser brightness (S(tot) = integrated fluxes) and maser luminosities (\Lup  (L$_{\odot}$); $L_{\rm p}$(photons s$^{-1}$)) are listed of the objects from Paper~I in 1987--2005 and from this paper in 1987--2015.  The definition of the levels (High, Mean, Low) is described in the text. Columns $\log$\,$L_{bol}$ and $\log$\,\mdot\ list stellar luminosities and mass-loss rates.}
\label{table:photon-luminosities}
\begin{center}
\begin{tabular}{lrr|c|rr|rr|rr|c|c}
\hline\noalign{\smallskip}
\multicolumn{1}{c}{Star} & \multicolumn{1}{c}{Type} & \multicolumn{1}{c|}{D} & log\,L$_{\rm H_2O}^{up}$ &
\multicolumn{6}{c|}{log\,S(tot) , log\,L$_p$} & log\,L$_{bol}$ & log\,\mdot \\[0.05cm]
\multicolumn{2}{c}{} & \multicolumn{1}{c|}{[pc]} & [L$_{\odot}$] &
\multicolumn{6}{c|}{[Jy~\kms] , [s$^{-1}$]} & [L$_{\odot}$]    &  [\Myr]   \\
&&&& \multicolumn{2}{c|}{High} & \multicolumn{2}{c|}{Mean}     & \multicolumn{2}{c|}{Low} &&\\
\hline\noalign{\smallskip}
R~Crt    & SRV &  236 & $-$4.88& 3.5 & 44.0& 2.9 & 43.4 &    2.3 &    42.8 &  $4.03\pm0.10$ & $-5.46$ \\ 
RT~Vir   & SRV &  226 & $-$5.16& 3.3 & 43.8& 2.8 & 43.3 &    2.3 &        42.8 &  $3.70\pm0.09$ & $-6.05$ \\ 
RX~Boo   & SRV &  136 & $-$6.23& 2.6 & 42.7& 1.9 & 41.9 &    1.5 &    41.5 &  $3.58\pm0.11$ & $-6.12$ \\ 
SV~Peg   & SRV &  333 & $-$6.42& 1.6 & 42.4& 0.9 & 41.7 & $<$0.8 & $<$41.6 &  $3.93\pm0.20$ & $-6.04$ \\[0.1cm]  
\noalign{\smallskip}\hline
\end{tabular}
\end{center}
Reference for distances: R~Crt, RT~Vir: see Table~\ref{centralcoords}; RX Boo: \cite{kamezaki12}; SV~Peg: \cite{sudou19}
\end{table*}

\subsection{\water\ maser, OH maser, and optical variability}
It is known from parallel monitoring programmes of Mira variables (\citealt{schwartz74}; \citealt{rudnitskii05} and references therein; Winnberg et al., in preparation) that the maser excitation responds to variations in the stellar optical light. Stars of type SRb, such as RX~Boo and SV~Peg (see paper I), R~Crt and RT~Vir, do not show large-amplitude periodic variations. The fluctuations of the maser spectral components seem to be so powerful that any responses of the \water\ maser emission to the small-amplitude variations of the optical light of SRb variable stars are effectively masked; in fact, we did not find any correlations between optical and \water\ maser emission for either of these three stars. This is not too different from the variations of main-line OH masers in SRbs, which were found in RT~Vir to have a similar radial distance from the star as \water\ masers \citep{bains03}. In R~Crt and RT~Vir, OH maser fluctuations and some periodicity was found, but this was apparently unrelated to any optical variations \citep{etoka01}. 

It should be noted that in SRVs of type SRa, such as the prototype W~Hya, correlations between optical and radio light curves have been detected (\citealt{rudnitskii99}; \citealt{imai19}). SRas have regular periodic optical variations like Mira variables, albeit with smaller amplitudes. As in Mira variables, the regular variations are sufficiently strong to be detected in the maser light curves even in the presence of irregular maser fluctuations.

\subsection{\water\ maser luminosities}
In Table~\ref{table:photon-luminosities}, we give luminosity information on R\,Crt and RT\,Vir, and on the two SRVs treated in Paper I (RX\,Boo, SV\,Peg). In column 4, we give \Lup, the potential maximum \water\ maser luminosity derived from the upper envelope. Assuming isotropy, \Lup\ = $4\pi D^2 \cdot S^{up}(tot)$ = $2.28\times10^{-14} \cdot S^{up}(tot) \cdot D^2$, with $S^{up}(tot)$ the integrated flux density in Jy\, \kms\ of the upper envelope spectrum and $D$ the distance in parsec (Col. 3); \Lup\ represents the maximum output which the source could produce if all the velocity components were to emit at their maximum level, at the same time and equally in all directions. The table also lists characteristic levels (high, mean, and low) of maser brightnesses, as given by integrated flux densities $S(tot)$ in Jy \kms\ (Cols. 5,7,9) and corresponding maser luminosities $L_{\rm p}$ in photons per second (Cols. 6,8,10). The photon luminosities were calculated assuming radial isotropy from the relation $L_{\rm p} = 4\pi D^2 \cdot S(tot) / E_{ph} = 6.0\times10^{35}  \cdot S(tot) \cdot D^2$, where $D$ is again the distance in parsec and $E_{ph} = h\nu_{22}$ is the energy of a photon at 22~GHz. The brightness of the mean level is the median of all integrated flux density measurements, while the high and low level is represented by the median of the seven highest and lowest integrated flux density measurements, respectively. This method of level calculation for the two latter levels was used to alleviate the dependency on individual measurements. The levels are representative for the years 1987-2015, except for RX~Boo and SV~Peg, where the integrated flux measurements were taken from Paper I for the years 1987-2005. 

Table~\ref{table:photon-luminosities} shows that the mean maser luminosities of R\,Crt and RT\,Vir are a factor of 25--50 higher than for the other two SRVs; the \Lup\ are 12--34 times higher. It is also shown that the ratio between the high and low level in each of the four SRVs is quite similar in the range $>$6--16 ($>$25--55 between \Lup\ and the low levels). Bursts in restricted velocity intervals in general do not influence total maser luminosities by factors of more than a few. In exceptional cases, this is different, as for the burst observed 1995 in RX~Boo (Paper I), which dominated the maser output for a few months. During this period, the maser luminosity was up to 50 times higher than the low level maser luminosity of RX~Boo. 

By its nature, \Lup\ represents the most reliable estimate of the potential \water\ maser emission because it eliminates the effects of the variability of the individual velocity components. Unfortunately the current sample is too small to take full advantage of this and potential correlations have not been brought to the fore.

To search for reasons of the systematic brightness differences between the SRVs, we compared the \water\ maser photon luminosities with the bolometric luminosities and mass-loss rates of the stars. The bolometric luminosities $L_{bol}$ (Col. 11) were determined from bolometric fluxes, as described in \cite{jimenez15}, and the distances listed in Table~\ref{table:photon-luminosities}. The mass-loss rates \mdot\ (Col. 12) were taken from \cite{olofsson02} scaled to the new distances used here. The results are listed in the last two columns of Table~\ref{table:photon-luminosities}. R~Crt has the highest stellar luminosity and mass-loss rate, but among the four SRVs, there is no evident correlation between the (mean) \water\ maser (photon) luminosity and either the stellar luminosity or the mass-loss rates. This could be due to either the small number of sources or the peculiarities of individual stars, or both. 

The RT~Vir maser may serve as example for peculiarity since the (mean) \water\ maser luminosity is comparable to R~Crt, although stellar luminosity, mass-loss rate, and also \Lup\  are two to four times lower than for R~Crt, indicating that the \water\ maser emission is exceptionally efficient. The stellar parameters of RT~Vir are comparing better with RX~Boo and SV~Peg, while their \water\ masers are significantly less luminous. 
Already \cite{bains03} noted the high \water\ maser luminosity of RT~Vir, on a level almost twice as luminous as the brightest \water\ masers of the Mira variables they studied. We conclude therefore that \water\ maser luminosities may differ by large factors between SRV's with otherwise similar stellar luminosities and mass-loss rates. This could be a temporary effect connected to mass-loss variations on timescales of tens of years, as we discussed before, when we considered the long-term variations of the brightness levels in individual stars. The alternative is that the masers are emitted non-isotropically, with RT~Vir's maser emitting preferentially into the line of sight. 

The \water\ maser luminosity distribution of SRVs determined by \cite{szymczak95} shows that the (mean) luminosities of all four SRVs discussed here are at the high end of the distribution. The \water\ maser luminosities of SRVs are generally a factor of 10--1000 weaker, with $\sim$80\% too low to be detected at levels of $\la 1$ Jy.

\subsection{Outflow velocities and velocity shifts}
The maximum outflow velocities reached by RT~Vir and R~Crt in their \water\ maser shells of $V_{\rm out}^{\rm max} = 9-10$~\kms\ are only marginally higher compared to RX~Boo and SV~Peg, with $V_{\rm out}^{\rm max} = 8\pm1$ \kms\ (Paper I), and the \water\ maser shell dimensions are also  similar. The boundaries vary with time, but for RT~Vir, RX~Boo, and R~Crt, they are found to be in the range of a few to $\sim$15 au for the inner rim  and 35--45 au for the outer rim. Based on average outflow velocities, the crossing times of matter through the \water\ maser shells are $\la$25 years. The maximum outflow velocities achieved in the \water\ maser shells of R~Crt and RX~Boo make up more than 85\% of the final expansion velocities of the stellar winds, and are already reached within a distance of $40\pm5$ au from the star. In RT~Vir, the  maximum outflow velocity reaches the final expansion velocity of the stellar wind already within the \water\ maser shell, at a distance of 15--27 au from the star. 

At any time, the \water\ maser emission is rarely detected over the maximum velocity range (cf. the FVt-plots; Figs. \ref{fig:rcrt-fvt} and \ref{fig:rtvir-fvt}). \cite{richards12} found emissivity variations across the maser shell with the brighter emission preferentially in the inner part of the shell. Assuming that the outflow velocities increase with the distance from the star, the emission from the more extreme parts of the velocity range are expected to be located close to the line of sight toward the central star in the outer parts of the shell.  \cite{richards12} also found that the filling factor of maser clouds in the shells is very small, so that at any time small variations in the number of clouds and in their emissivity will influence the detectability of maser emission at the extreme velocities. The outer edges of velocity profiles, as seen in the FVt-plots, indicate that the radial emissivity pattern is different for the front and rear parts of the maser shells of R~Crt and RT~Vir, and will likely change over time. Relatively strong emission is seen for both stars at the blueshifted border of the velocity profile belonging to the outer part of the  maser shell located in front of the stars. The opposite is true for the redshifted border. This results in rather weakly variable blue velocity borders, while the red velocity borders show a ragged appearance. We anticipate that this behaviour can change on the time scales of the crossing times through the \water\ maser shell, when stronger emission clouds appear or disappear close to the line of  sight.

Occasionally \water\ maser emission is detected with velocities larger than the expansion velocities $V_{\rm exp}$ given by CO and OH maser emission. This is reported for R~Crt, RT~Vir \citep{bowers94}, and RX~Boo (Paper I), but not further discussed in these works because the overshoot is seen only in individual observations, with an excess velocity of at most a few \kms and the uncertainty of $V_{\rm exp}$ was often of the same order. The transient overshoot might, however,  prove to be a real effect and would provide evidence for the presence of turbulence-induced velocity fluctuations or shock waves in the stellar wind. 

The spectral features in the maser profiles of R~Crt and RT~Vir are so numerous that they are generally blended in velocity space. With Gaussian fits to the features, this was previously found by \cite{lekht99} for RT~Vir. We also found a prevalence of blending in similar stars, such as RX~Boo (Paper I), and the Mira variable U~Her (Winnberg et al., in preparation), for which we made a similar line fit analysis. 

It is, therefore, difficult to measure velocity shifts in the spectra, especially for neighbouring components of similar flux density. Brightness fluctuations lead to shifts of the velocity centre of the blended line profile, which can mask the systematic shifts expected when the maser cloud moves to higher velocity with time. It is necessary to spatially resolve the maser clouds in such cases to remove the ambiguity (see \citealt{imai97b}). An exceptional case is the 11~\kms\ feature in RT~Vir, in which no velocity shift $\ge 0.06$ \kms~yr$^{-1}$ was detected in the absence of blending from neighbouring features over 7.5 years. 

\subsection{Limitations of the standard model of the CSE \label{cse-model}}
The properties of the \water\ maser in evolved stars are usually discussed within the framework of the standard CSE model (\citealt{hoefner18}; Sect.~\ref{presdata}). In the case of RT~Vir, this approach is explicitly supported by the 3D kinematics presented by \cite{imai03}. Also, the coincidence ($\le 0.4$ \kms) of the centre of the observed \water\ maser velocity range with the stellar radial velocity in both stars is as predicted from the model. It allows for the placement of the unknown position of the star close to the centre of maps of the spatial distribution of the maser components. 

However,  the presence of asymmetric structures in the \water\ maser region are often suspected, as, for example, a bipolar flow in the case of R~Crt \citep{ishitsuka01}. Ishitsuka et al. found the blueshifted emission in 1998 mostly in the south-east of the maser distribution and the redshifted emission in the north-west, with the centre of expansion (identified as the star) in the middle. Due to the transient nature of the masers, such structures are difficult to confirm with measurements that have been taken several years apart. In the case of R~Crt, the presence of an asymmetric structure could, however, be confirmed by \cite{khouri20}, who found a biconical structure in the polarised emission of dust with about the same position angle on the sky as the \water\ maser bipolar flow. 

If asymmetric structures determine the (preferred) location of the \water\ maser emission regions in the stellar surroundings, then the preference of enhanced maser emission in particular velocity intervals found in the long-term variability pattern of R~Crt and RT~Vir could easily be explained. Likewise, the assumption of isotropic emission used to compare \water\ maser luminosities with stellar luminosities and mass-loss rates could be invalidated and the relatively high water maser luminosity of RT~Vir might then be due to a chance coincidence of emission larger than average into our line of sight. The asymmetric structures may be an additional component to the underlying framework of a spherically symmetric expanding shell. Their origin might be directional differences in the mass-loss due to convection cells (\citealt{freytag17}; \citealt{perrin20}) or deviations from spherical mass loss due to the presence of a companion \citep{chen17}. We expect that the long-living patterns in the \water\ maser profiles should change on the timescales of the shell-crossing times ($\la$25 years).

\begin{figure*}
\resizebox{18cm}{!}{
\includegraphics{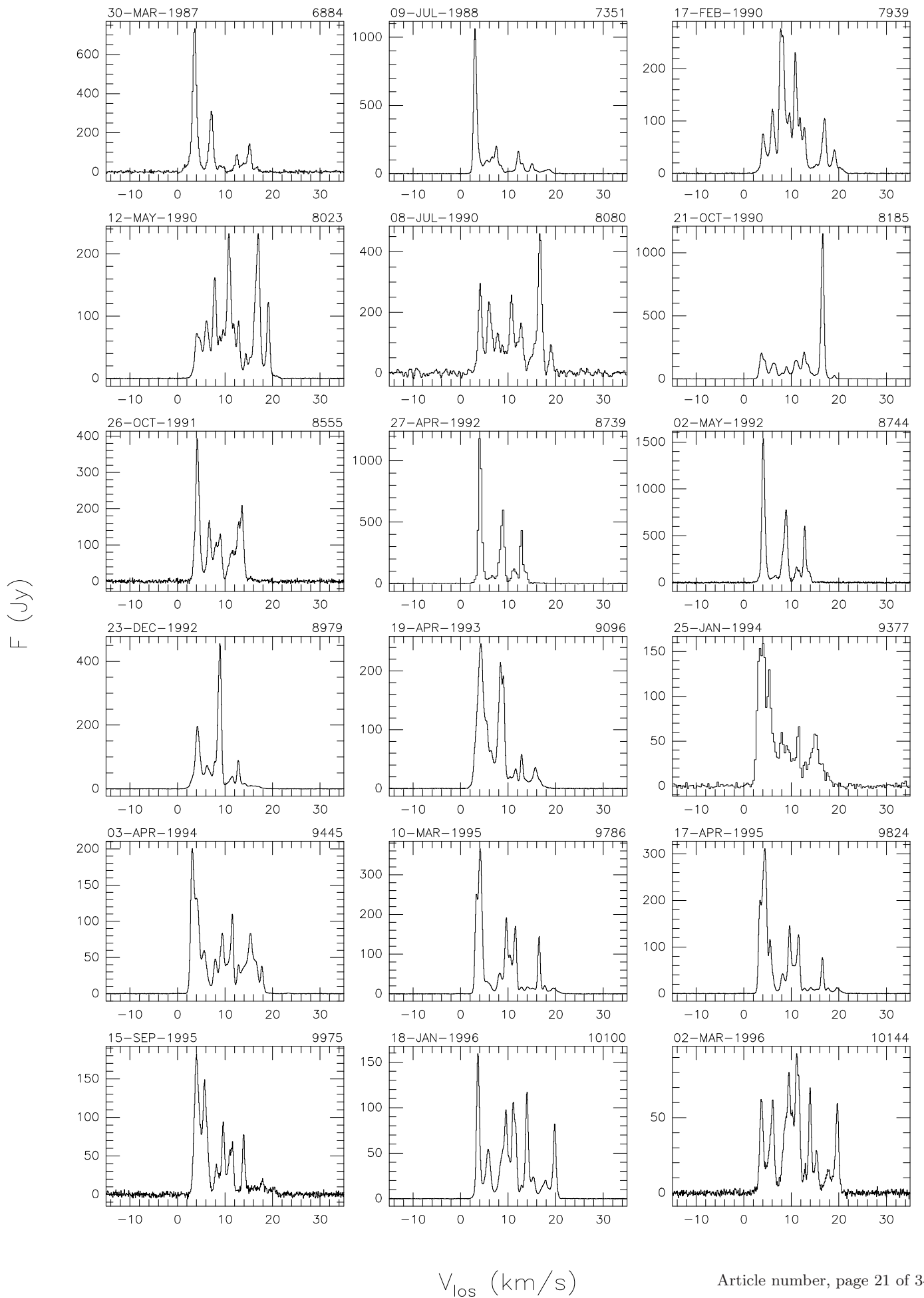}}
\caption{All H$_2$O maser spectra of R~Crt. The observing date (top left) and TJD (top right) are indicated for each spectrum.}
\label{fig:rcrt_all}
\end{figure*}

The validity of the standard CSE model for the interpretation of the \water\ maser properties is put to test by the simultaneous observations of SiO and \water\ masers, with the position of the SiO maser as indicator for the stellar position \citep{dodson14}. In R~Crt, \cite{kim18} measured, in 2015 and 2016, a one-sided displacement of the \water\ maser distribution with respect to the SiO masers, which apparently indicates a strong asymmetry of the maser excitation in the environment of the star and is in contradiction to the location of the star by \cite{ishitsuka01}. 
At the dates of the observations reported by \cite{kim18}, the emission at redshifted velocities $V>14$ \kms\ was relatively weak (see also our spectra taken in 2015 and shown in Fig. \ref{fig:rcrt_all}). The redshifted emission may not have been detected, explaining the apparent one-sided displacement and alleviating the apparent contradiction between the results of \cite{ishitsuka01} and \cite{kim18}.

\section{Conclusions \label{conclusions}} 
The predominant type of \water\ maser variability in R~Crt and
RT~Vir are short-term brightness variations of individual spectral components on timescales of months to up to 1.5 years. These maser fluctuations are superimposed on long-term variations of the integrated flux density on much longer timescales. They are caused by brightness variations, in particular velocity intervals which, independently from one another, determine the level of the integrated flux density on timescales of the order of a decade and more. Both stars are SRVs of optical variability type SRb, as in the case of RX~Boo and SV~Peg (Paper I). The optical variability is of small amplitude and mostly is not regular. If these variations have influence on the \water\ maser excitation conditions, they are of minor importance since we have not found any correlation between \water\ maser and optical light curves.

Velocity shifts of individual spectral components were observed but, in general, we could not distinguish between real shifts due to the movement of an emitting maser cloud with the stellar wind or shifts induced by blends of maser lines with velocities close to each other. One exception was the 11 \kms\ feature of RT~Vir, where the absence of any velocity shift could be measured over 7.5 years, while this feature was much brighter than the features in its vicinity.

While the short-term variations are due to a coming and going of individual maser clouds, the long-term brightness changes are probably due to the movement through the \water\ maser shell of larger regions, in which masers are preferentially excited compared to other parts of the shell. We speculate that they have higher densities and host more potentially excitable maser clouds than are present on average in the shell. These regions typically would need two to three decades  to cross the \water\ maser shell, which would provide a natural explanation of the longer timescales observed. We conclude from the absence of a velocity shift of the 11 \kms\ feature  that the emitting cloud was moving at radii where in RT~Vir's shell the terminal expansion velocity had already been reached.

The brightness variations imply that also the maser luminosities are strongly varying. There is typically a difference of a factor of ten between low and high states of maser activity. The mean maser luminosities of R~Crt and RT~Vir are an order of magnitude larger than those of RX~Boo and SV~Peg, but this is not correlated with stellar luminosity or mass-loss rates. The mean luminosities are averages of 10--20 years, but because of the long-term brightness variations, even these averages themselves are valid only for a given period of time and may not necessarily be representative for other epochs.

The interferometric maps in which the observed maser distributions are in conflict with the standard model of a spherically symmetric expanding shell indicate the presence of inhomogeneities in the stellar wind. The presence of long-living structures corroborates this. Monitoring the strong \water\ maser emission allows us to probe the dynamical processes in the CSE of semi-regular variable stars at radial distances from a few au up to $\sim$45~au. A detailed study requires frequent observations with single-dish telescopes with cadences of a month or less to monitor the short-term brightness fluctuations, along with regular revisits every few years to monitor the long-term asymmetries in the variability pattern. 

\begin{acknowledgements}
The Medicina 32-m data presented here are part of a long-term monitoring programme, which concerned both evolved stars and star-forming regions. The paper on water masers in star-forming regions was published in 2007 (Felli et al.), and most of the people on that paper also helped with the observations of the stellar masers presented here. We thank those who did. In particular we are grateful to Gianni Comoretto and Riccardo Cesaroni (INAF-Oss. Astrof. Arcetri) for building the auto-correlator and fixing it when necessary, and for writing the routines that allow us to present the data in an elegant manner, respectively.
We are grateful to the staff at Medicina observatory for their expert assistance and technical problem-solving. It was Malcolm Walmsley who first made the suggestion that it would be useful to monitor water masers with the Medicina and Effelsberg single-dish antennas, and to do multi-epoch interferometric observations of the targets from time to time. For data reduction and the preparation of figures GILDAS software available at
\mbox {www.iram.fr/IRAMFR/GILDAS} was used.  
This research has made use of the SIMBAD database, operated at CDS, Strasbourg, France, and of NASA's Astrophysics Data System. To prepare this paper we used the LaTeX editor Overleaf.
We acknowledge with thanks the variable star observations from the AAVSO International Database contributed by observers worldwide and used in this research. 
\end{acknowledgements}



\bibliographystyle{aa}
\bibliography{masterlist-biblio.bib}

\begin{thebibliography}{63}
\expandafter\ifx\csname natexlab\endcsname\relax\def\natexlab#1{#1}\fi

\bibitem[{{Bains} {et~al.}(2003){Bains}, {Cohen}, {Louridas}, {Richards},
  {Rosa-Gonz{\'a}lez}, \& {Yates}}]{bains03}
{Bains}, I., {Cohen}, R.~J., {Louridas}, A., {et~al.} 2003, \mnras, 342, 8

\bibitem[{{Barvainis} \& {Deguchi}(1989)}]{barvainis89}
{Barvainis}, R. \& {Deguchi}, S. 1989, \aj, 97, 1089

\bibitem[{{Berulis} {et~al.}(1987){Berulis}, {Lekht}, \&
  {Pashchenko}}]{berulis87}
{Berulis}, I.~I., {Lekht}, E.~E., \& {Pashchenko}, M.~I. 1987, Soviet Astronomy
  Letters, 13, 124

\bibitem[{{Berulis} {et~al.}(1983){Berulis}, {Lekht}, {Pashchenko}, \&
  {Rudnitskii}}]{berulis83}
{Berulis}, I.~I., {Lekht}, E.~E., {Pashchenko}, M.~I., \& {Rudnitskii}, G.~M.
  1983, \sovast, 27, 179

\bibitem[{{Bowers} {et~al.}(1993){Bowers}, {Claussen}, \&
  {Johnston}}]{bowers93}
{Bowers}, P.~F., {Claussen}, M.~J., \& {Johnston}, K.~J. 1993, \aj, 105, 284

\bibitem[{{Bowers} \& {Johnston}(1994)}]{bowers94}
{Bowers}, P.~F. \& {Johnston}, K.~J. 1994, \apjs, 92, 189

\bibitem[{{Cernicharo} {et~al.}(1997){Cernicharo}, {Alcolea}, {Baudry}, \&
  {Gonzalez-Alfonso}}]{cernicharo97}
{Cernicharo}, J., {Alcolea}, J., {Baudry}, A., \& {Gonzalez-Alfonso}, E. 1997,
  \aap, 319, 607

\bibitem[{{Chen} {et~al.}(2017){Chen}, {Frank}, {Blackman}, {Nordhaus}, \&
  {Carroll-Nellenback}}]{chen17}
{Chen}, Z., {Frank}, A., {Blackman}, E.~G., {Nordhaus}, J., \&
  {Carroll-Nellenback}, J. 2017, \mnras, 468, 4465

\bibitem[{{Colomer} {et~al.}(2000){Colomer}, {Reid}, {Menten}, \&
  {Bujarrabal}}]{colomer00}
{Colomer}, F., {Reid}, M.~J., {Menten}, K.~M., \& {Bujarrabal}, V. 2000, \aap,
  355, 979

\bibitem[{{Comoretto} {et~al.}(1990){Comoretto}, {Palagi}, {Cesaroni}, {Felli},
  {Bettarini}, {Catarzi}, {Curioni}, {Curioni}, {Di Franco}, {Giovanardi},
  {Massi}, {Palla}, {Panella}, {Rossi}, {Speroni}, \& {Tofani}}]{comoretto90}
{Comoretto}, G., {Palagi}, F., {Cesaroni}, R., {et~al.} 1990, \aaps, 84, 179

\bibitem[{{Cooke} \& {Elitzur}(1985)}]{cooke85}
{Cooke}, B. \& {Elitzur}, M. 1985, \apj, 295, 175

\bibitem[{{Cox} \& {Parker}(1979)}]{cox79}
{Cox}, G.~G. \& {Parker}, E.~A. 1979, \mnras, 186, 197

\bibitem[{{D{\'\i}az-Luis} {et~al.}(2019){D{\'\i}az-Luis}, {Alcolea},
  {Bujarrabal}, {Santand er-Garc{\'\i}a}, {Castro-Carrizo},
  {G{\'o}mez-Garrido}, \& {Desmurs}}]{diazluis19}
{D{\'\i}az-Luis}, J.~J., {Alcolea}, J., {Bujarrabal}, V., {et~al.} 2019, \aap,
  629, A94

\bibitem[{{Dickinson}(1976)}]{dickinson76}
{Dickinson}, D.~F. 1976, \apjs, 30, 259

\bibitem[{{Dickinson} {et~al.}(1973){Dickinson}, {Bechis}, \&
  {Barrett}}]{dickinson73}
{Dickinson}, D.~F., {Bechis}, K.~P., \& {Barrett}, A.~H. 1973, \apj, 180, 831

\bibitem[{{Dodson} {et~al.}(2014){Dodson}, {Rioja}, {Jung}, {Sohn}, {Byun},
  {Cho}, {Lee}, {Kim}, {Kim}, {Oh}, {Han}, {Je}, {Chung}, {Wi}, {Kang}, {Lee},
  {Chung}, {Kim}, {Kim}, {Lee}, {Roh}, {Oh}, {Yeom}, {Song}, \&
  {Kang}}]{dodson14}
{Dodson}, R., {Rioja}, M.~J., {Jung}, T.-H., {et~al.} 2014, \aj, 148, 97

\bibitem[{{Engels} {et~al.}(1988){Engels}, {Schmid-Burgk}, \&
  {Walmsley}}]{engels88}
{Engels}, D., {Schmid-Burgk}, J., \& {Walmsley}, C.~M. 1988, \aap, 191, 283

\bibitem[{{Engels} {et~al.}(1997){Engels}, {Winnberg}, {Walmsley}, \&
  {Brand}}]{engels97}
{Engels}, D., {Winnberg}, A., {Walmsley}, C.~M., \& {Brand}, J. 1997, \aap,
  322, 291

\bibitem[{{Etoka} {et~al.}(2001){Etoka}, {B{\l}aszkiewicz}, {Szymczak}, \& {Le
  Squeren}}]{etoka01}
{Etoka}, S., {B{\l}aszkiewicz}, L., {Szymczak}, M., \& {Le Squeren}, A.~M.
  2001, \aap, 378, 522

\bibitem[{{Felli} {et~al.}(2007){Felli}, {Brand}, {Cesaroni}, {Codella},
  {Comoretto}, {Di Franco}, {Massi}, {Moscadelli}, {Nesti}, {Olmi}, {Palagi},
  {Panella}, \& {Valdettaro}}]{felli07}
{Felli}, M., {Brand}, J., {Cesaroni}, R., {et~al.} 2007, \aap, 476, 373

\bibitem[{{Freytag} {et~al.}(2017){Freytag}, {Liljegren}, \&
  {H{\"o}fner}}]{freytag17}
{Freytag}, B., {Liljegren}, S., \& {H{\"o}fner}, S. 2017, \aap, 600, A137

\bibitem[{{Gaia Collaboration} {et~al.}(2018){Gaia Collaboration}, {Brown},
  {Vallenari}, {Prusti}, {de Bruijne}, {Babusiaux}, {Bailer-Jones}, {Biermann},
  {Evans}, {Eyer}, {Jansen}, {Jordi}, {Klioner}, {Lammers}, {Lindegren},
  {Luri}, {Mignard}, {Panem}, {Pourbaix}, {Randich}, {Sartoretti}, {Siddiqui},
  {Soubiran}, {van Leeuwen}, {Walton}, {Arenou}, {Bastian}, {Cropper},
  {Drimmel}, {Katz}, {Lattanzi}, {Bakker}, {Cacciari}, {Casta{\~n}eda},
  {Chaoul}, {Cheek}, {De Angeli}, {Fabricius}, {Guerra}, {Holl}, {Masana},
  {Messineo}, {Mowlavi}, {Nienartowicz}, {Panuzzo}, {Portell}, {Riello},
  {Seabroke}, {Tanga}, {Th{\'e}venin}, {Gracia-Abril}, {Comoretto},
  {Garcia-Reinaldos}, {Teyssier}, {Altmann}, {Andrae}, {Audard},
  {Bellas-Velidis}, {Benson}, {Berthier}, {Blomme}, {Burgess}, {Busso},
  {Carry}, {Cellino}, {Clementini}, {Clotet}, {Creevey}, {Davidson}, {De
  Ridder}, {Delchambre}, {Dell'Oro}, {Ducourant},
  {Fern{\'a}ndez-Hern{\'a}ndez}, {Fouesneau}, {Fr{\'e}mat}, {Galluccio},
  {Garc{\'\i}a-Torres}, {Gonz{\'a}lez-N{\'u}{\~n}ez}, {Gonz{\'a}lez-Vidal},
  {Gosset}, {Guy}, {Halbwachs}, {Hambly}, {Harrison}, {Hern{\'a}ndez},
  {Hestroffer}, {Hodgkin}, {Hutton}, {Jasniewicz}, {Jean-Antoine-Piccolo},
  {Jordan}, {Korn}, {Krone-Martins}, {Lanzafame}, {Lebzelter}, {L{\"o}ffler},
  {Manteiga}, {Marrese}, {Mart{\'\i}n-Fleitas}, {Moitinho}, {Mora}, {Muinonen},
  {Osinde}, {Pancino}, {Pauwels}, {Petit}, {Recio-Blanco}, {Richards},
  {Rimoldini}, {Robin}, {Sarro}, {Siopis}, {Smith}, {Sozzetti}, {S{\"u}veges},
  {Torra}, {van Reeven}, {Abbas}, {Abreu Aramburu}, {Accart}, {Aerts},
  {Altavilla}, {{\'A}lvarez}, {Alvarez}, {Alves}, {Anderson}, {Andrei},
  {Anglada Varela}, {Antiche}, {Antoja}, {Arcay}, {Astraatmadja}, {Bach},
  {Baker}, {Balaguer-N{\'u}{\~n}ez}, {Balm}, {Barache}, {Barata}, {Barbato},
  {Barblan}, {Barklem}, {Barrado}, {Barros}, {Barstow}, {Bartholom{\'e}
  Mu{\~n}oz}, {Bassilana}, {Becciani}, {Bellazzini}, {Berihuete}, {Bertone},
  {Bianchi}, {Bienaym{\'e}}, {Blanco-Cuaresma}, {Boch}, {Boeche}, {Bombrun},
  {Borrachero}, {Bossini}, {Bouquillon}, {Bourda}, {Bragaglia}, {Bramante},
  {Breddels}, {Bressan}, {Brouillet}, {Br{\"u}semeister}, {Brugaletta},
  {Bucciarelli}, {Burlacu}, {Busonero}, {Butkevich}, {Buzzi}, {Caffau},
  {Cancelliere}, {Cannizzaro}, {Cantat-Gaudin}, {Carballo}, {Carlucci},
  {Carrasco}, {Casamiquela}, {Castellani}, {Castro-Ginard}, {Charlot},
  {Chemin}, {Chiavassa}, {Cocozza}, {Costigan}, {Cowell}, {Crifo}, {Crosta},
  {Crowley}, {Cuypers}, {Dafonte}, {Damerdji}, {Dapergolas}, {David}, {David},
  {de Laverny}, {De Luise}, {De March}, {de Martino}, {de Souza}, {de Torres},
  {Debosscher}, {del Pozo}, {Delbo}, {Delgado}, {Delgado}, {Di Matteo},
  {Diakite}, {Diener}, {Distefano}, {Dolding}, {Drazinos}, {Dur{\'a}n},
  {Edvardsson}, {Enke}, {Eriksson}, {Esquej}, {Eynard Bontemps}, {Fabre},
  {Fabrizio}, {Faigler}, {Falc{\~a}o}, {Farr{\`a}s Casas}, {Federici},
  {Fedorets}, {Fernique}, {Figueras}, {Filippi}, {Findeisen}, {Fonti},
  {Fraile}, {Fraser}, {Fr{\'e}zouls}, {Gai}, {Galleti}, {Garabato},
  {Garc{\'\i}a-Sedano}, {Garofalo}, {Garralda}, {Gavel}, {Gavras}, {Gerssen},
  {Geyer}, {Giacobbe}, {Gilmore}, {Girona}, {Giuffrida}, {Glass}, {Gomes},
  {Granvik}, {Gueguen}, {Guerrier}, {Guiraud}, {Guti{\'e}rrez-S{\'a}nchez},
  {Haigron}, {Hatzidimitriou}, {Hauser}, {Haywood}, {Heiter}, {Helmi}, {Heu},
  {Hilger}, {Hobbs}, {Hofmann}, {Holland}, {Huckle}, {Hypki}, {Icardi},
  {Jan{\ss}en}, {Jevardat de Fombelle}, {Jonker}, {Juh{\'a}sz}, {Julbe},
  {Karampelas}, {Kewley}, {Klar}, {Kochoska}, {Kohley}, {Kolenberg},
  {Kontizas}, {Kontizas}, {Koposov}, {Kordopatis}, {Kostrzewa-Rutkowska},
  {Koubsky}, {Lambert}, {Lanza}, {Lasne}, {Lavigne}, {Le Fustec}, {Le
  Poncin-Lafitte}, {Lebreton}, {Leccia}, {Leclerc}, {Lecoeur-Taibi},
  {Lenhardt}, {Leroux}, {Liao}, {Licata}, {Lindstr{\o}m}, {Lister}, {Livanou},
  {Lobel}, {L{\'o}pez}, {Managau}, {Mann}, {Mantelet}, {Marchal}, {Marchant},
  {Marconi}, {Marinoni}, {Marschalk{\'o}}, {Marshall}, {Martino}, {Marton},
  {Mary}, {Massari}, {Matijevi{\v{c}}}, {Mazeh}, {McMillan}, {Messina},
  {Michalik}, {Millar}, {Molina}, {Molinaro}, {Moln{\'a}r}, {Montegriffo},
  {Mor}, {Morbidelli}, {Morel}, {Morris}, {Mulone}, {Muraveva}, {Musella},
  {Nelemans}, {Nicastro}, {Noval}, {O'Mullane}, {Ord{\'e}novic},
  {Ord{\'o}{\~n}ez-Blanco}, {Osborne}, {Pagani}, {Pagano}, {Pailler},
  {Palacin}, {Palaversa}, {Panahi}, {Pawlak}, {Piersimoni}, {Pineau}, {Plachy},
  {Plum}, {Poggio}, {Poujoulet}, {Pr{\v{s}}a}, {Pulone}, {Racero}, {Ragaini},
  {Rambaux}, {Ramos-Lerate}, {Regibo}, {Reyl{\'e}}, {Riclet}, {Ripepi}, {Riva},
  {Rivard}, {Rixon}, {Roegiers}, {Roelens}, {Romero-G{\'o}mez}, {Rowell},
  {Royer}, {Ruiz-Dern}, {Sadowski}, {Sagrist{\`a} Sell{\'e}s}, {Sahlmann},
  {Salgado}, {Salguero}, {Sanna}, {Santana-Ros}, {Sarasso}, {Savietto},
  {Schultheis}, {Sciacca}, {Segol}, {Segovia}, {S{\'e}gransan}, {Shih},
  {Siltala}, {Silva}, {Smart}, {Smith}, {Solano}, {Solitro}, {Sordo}, {Soria
  Nieto}, {Souchay}, {Spagna}, {Spoto}, {Stampa}, {Steele},
  {Steidelm{\"u}ller}, {Stephenson}, {Stoev}, {Suess}, {Surdej}, {Szabados},
  {Szegedi-Elek}, {Tapiador}, {Taris}, {Tauran}, {Taylor}, {Teixeira},
  {Terrett}, {Teyssand ier}, {Thuillot}, {Titarenko}, {Torra Clotet}, {Turon},
  {Ulla}, {Utrilla}, {Uzzi}, {Vaillant}, {Valentini}, {Valette}, {van Elteren},
  {Van Hemelryck}, {van Leeuwen}, {Vaschetto}, {Vecchiato}, {Veljanoski},
  {Viala}, {Vicente}, {Vogt}, {von Essen}, {Voss}, {Votruba}, {Voutsinas},
  {Walmsley}, {Weiler}, {Wertz}, {Wevers}, {Wyrzykowski}, {Yoldas},
  {{\v{Z}}erjal}, {Ziaeepour}, {Zorec}, {Zschocke}, {Zucker}, {Zurbach}, \&
  {Zwitter}}]{gaiacoll18}
{Gaia Collaboration}, {Brown}, A.~G.~A., {Vallenari}, A., {et~al.} 2018, \aap,
  616, A1

\bibitem[{{Goldreich} \& {Scoville}(1976)}]{goldreich76}
{Goldreich}, P. \& {Scoville}, N. 1976, \apj, 205, 144

\bibitem[{{Gomez Balboa} \& {L{\'e}pine}(1986)}]{gomezbalboa86}
{Gomez Balboa}, A.~M. \& {L{\'e}pine}, J.~R.~D. 1986, \aap, 159, 166

\bibitem[{{H{\"o}fner} \& {Olofsson}(2018)}]{hoefner18}
{H{\"o}fner}, S. \& {Olofsson}, H. 2018, \aapr, 26, 1

\bibitem[{{Imai} {et~al.}(2019){Imai}, {Nakagawa}, \& {Takaba}}]{imai19}
{Imai}, H., {Nakagawa}, A., \& {Takaba}, H. 2019, \pasj, 71, 120

\bibitem[{{Imai} {et~al.}(1997{\natexlab{a}}){Imai}, {Sasao}, {Kameya},
  {Miyoshi}, {Shibata}, {Asaki}, {Omodaka}, {Morimoto}, {Mochizuki},
  {Suzuyama}, {Iguchi}, {Kameno}, {Jike}, {Iwadate}, {Sakai}, {Miyaji},
  {Kawaguchi}, \& {Miyazawa}}]{imai97a}
{Imai}, H., {Sasao}, T., {Kameya}, O., {et~al.} 1997{\natexlab{a}}, \aap, 317,
  L67

\bibitem[{{Imai} {et~al.}(2003){Imai}, {Shibata}, {Marvel}, {Diamond}, {Sasao},
  {Miyoshi}, {Inoue}, {Migenes}, \& {Murata}}]{imai03}
{Imai}, H., {Shibata}, K.~M., {Marvel}, K.~B., {et~al.} 2003, \apj, 590, 460

\bibitem[{{Imai} {et~al.}(1997{\natexlab{b}}){Imai}, {Shibata}, {Sasao},
  {Miyoshi}, {Kameya}, {Omodaka}, {Morimoto}, {Iwata}, {Suzuyama}, {Mochizuki},
  {Miyaji}, \& {Takeuti}}]{imai97b}
{Imai}, H., {Shibata}, K.~M., {Sasao}, T., {et~al.} 1997{\natexlab{b}}, \aap,
  319, L1

\bibitem[{{Ishitsuka} {et~al.}(2001){Ishitsuka}, {Imai}, {Omodaka}, {Ueno},
  {Kameya}, {Sasao}, {Morimoto}, {Miyaji}, {Nakajima}, \&
  {Watanabe}}]{ishitsuka01}
{Ishitsuka}, J.~K., {Imai}, H., {Omodaka}, T., {et~al.} 2001, \pasj, 53, 1231

\bibitem[{{Jim{\'e}nez-Esteban} \& {Engels}(2015)}]{jimenez15}
{Jim{\'e}nez-Esteban}, F.~M. \& {Engels}, D. 2015, \aap, 579, A76

\bibitem[{{Johnston} {et~al.}(1985){Johnston}, {Spencer}, \&
  {Bowers}}]{johnston85}
{Johnston}, K.~J., {Spencer}, J.~H., \& {Bowers}, P.~F. 1985, \apj, 290, 660

\bibitem[{{Kamezaki} {et~al.}(2012){Kamezaki}, {Nakagawa}, {Omodaka},
  {Kurayama}, {Imai}, {Tafoya}, {Matsui}, {Nishida}, {Nagayama}, {Honma},
  {Kobayashi}, {Miyaji}, \& {Mine}}]{kamezaki12}
{Kamezaki}, T., {Nakagawa}, A., {Omodaka}, T., {et~al.} 2012, \pasj, 64, 7

\bibitem[{{Kerschbaum} \& {Olofsson}(1999)}]{kerschbaum99}
{Kerschbaum}, F. \& {Olofsson}, H. 1999, \aaps, 138, 299

\bibitem[{{Khouri} {et~al.}(2020){Khouri}, {Vlemmings}, {Paladini}, {Ginski},
  {Lagadec}, {Maercker}, {Kervella}, {De Beck}, {Decin}, {de Koter}, \&
  {Waters}}]{khouri20}
{Khouri}, T., {Vlemmings}, W.~H.~T., {Paladini}, C., {et~al.} 2020, \aap, 635,
  A200

\bibitem[{{Kim} {et~al.}(2018){Kim}, {Cho}, {Yun}, {Choi}, {Yoon}, {Kim},
  {Dodson}, {Rioja}, {Yang}, \& {Yoon}}]{kim18}
{Kim}, D.-J., {Cho}, S.-H., {Yun}, Y., {et~al.} 2018, \apjl, 866, L19

\bibitem[{{Kim} {et~al.}(2010){Kim}, {Cho}, {Oh}, \& {Byun}}]{kim10}
{Kim}, J., {Cho}, S.-H., {Oh}, C.~S., \& {Byun}, D.-Y. 2010, \apjs, 188, 209

\bibitem[{{Leal-Ferreira} {et~al.}(2013){Leal-Ferreira}, {Vlemmings},
  {Kemball}, \& {Amiri}}]{lealferreira13}
{Leal-Ferreira}, M.~L., {Vlemmings}, W.~H.~T., {Kemball}, A., \& {Amiri}, N.
  2013, \aap, 554, A134

\bibitem[{{Lekht} {et~al.}(1999){Lekht}, {Mendoza-Torres}, {Pashchenko}, \&
  {Berulis}}]{lekht99}
{Lekht}, E.~E., {Mendoza-Torres}, J.~E., {Pashchenko}, M.~I., \& {Berulis},
  I.~I. 1999, \aap, 343, 241

\bibitem[{{Likkel} {et~al.}(1992){Likkel}, {Morris}, \& {Maddalena}}]{likkel92}
{Likkel}, L., {Morris}, M., \& {Maddalena}, R.~J. 1992, \aap, 256, 581

\bibitem[{{Mendoza-Torres} {et~al.}(1997){Mendoza-Torres}, {Lekht}, {Berulis},
  \& {Pashchenko}}]{mendozatorres97}
{Mendoza-Torres}, J.~E., {Lekht}, E.~E., {Berulis}, I.~I., \& {Pashchenko},
  M.~I. 1997, \aaps, 126, 257

\bibitem[{{Migenes} {et~al.}(1999){Migenes}, {Horiuchi}, {Slysh}, {Val'TTS},
  {Golubev}, {Edwards}, {Fomalont}, {Okayasu}, {Diamond}, {Umemoto}, {Shibata},
  \& {Inoue}}]{migenes99}
{Migenes}, V., {Horiuchi}, S., {Slysh}, V.~I., {et~al.} 1999, \apjs, 123, 487

\bibitem[{{Murakawa} {et~al.}(2003){Murakawa}, {Yates}, {Richards}, \&
  {Cohen}}]{murakawa03}
{Murakawa}, K., {Yates}, J.~A., {Richards}, A.~M.~S., \& {Cohen}, R.~J. 2003,
  \mnras, 344, 1

\bibitem[{{Olofsson} {et~al.}(2002){Olofsson}, {Gonz{\'a}lez Delgado},
  {Kerschbaum}, \& {Sch{\"o}ier}}]{olofsson02}
{Olofsson}, H., {Gonz{\'a}lez Delgado}, D., {Kerschbaum}, F., \& {Sch{\"o}ier},
  F.~L. 2002, \aap, 391, 1053

\bibitem[{{Ott} {et~al.}(1994){Ott}, {Witzel}, {Quirrenbach}, {Krichbaum},
  {Standke}, {Schalinski}, \& {Hummel}}]{ott94}
{Ott}, M., {Witzel}, A., {Quirrenbach}, A., {et~al.} 1994, \aap, 284, 331

\bibitem[{{Perrin} {et~al.}(2020){Perrin}, {Ridgway}, {Lacour}, {Haubois},
  {Thiebaut}, {Berger}, {Lacasse}, {Millan-Gabet}, {Monnier}, {Pedretti},
  {Ragland }, \& {Traub}}]{perrin20}
{Perrin}, G., {Ridgway}, S.~T., {Lacour}, S., {et~al.} 2020, arXiv e-prints,
  arXiv:2008.09801

\bibitem[{{Reid} \& {Menten}(1990)}]{reid90}
{Reid}, M.~J. \& {Menten}, K.~M. 1990, \apjl, 360, L51

\bibitem[{{Richards} {et~al.}(2012){Richards}, {Etoka}, {Gray}, {Lekht},
  {Mendoza-Torres}, {Murakawa}, {Rudnitskij}, \& {Yates}}]{richards12}
{Richards}, A.~M.~S., {Etoka}, S., {Gray}, M.~D., {et~al.} 2012, \aap, 546, A16

\bibitem[{{Rudnitskii} {et~al.}(1999){Rudnitskii}, {Lekht}, \&
  {Berulis}}]{rudnitskii99}
{Rudnitskii}, G.~M., {Lekht}, E.~E., \& {Berulis}, I.~I. 1999, Astronomy
  Letters, 25, 398

\bibitem[{{Rudnitskii} \& {Pashchenko}(2005)}]{rudnitskii05}
{Rudnitskii}, G.~M. \& {Pashchenko}, M.~I. 2005, Astronomy Letters, 31, 760

\bibitem[{{Samus'} {et~al.}(2017){Samus'}, {Kazarovets}, {Durlevich},
  {Kireeva}, \& {Pastukhova}}]{samus17}
{Samus'}, N.~N., {Kazarovets}, E.~V., {Durlevich}, O.~V., {Kireeva}, N.~N., \&
  {Pastukhova}, E.~N. 2017, Astronomy Reports, 61, 80

\bibitem[{{Schwartz} {et~al.}(1974){Schwartz}, {Harvey}, \&
  {Barrett}}]{schwartz74}
{Schwartz}, P.~R., {Harvey}, P.~M., \& {Barrett}, A.~H. 1974, \apj, 187, 491

\bibitem[{{Shintani} {et~al.}(2008){Shintani}, {Imai}, {Ando}, {Nakashima},
  {Hirota}, {Inomata}, {Kai}, {Kameno}, {Kijima}, {Kobayashi}, {Kuroki},
  {Maeda}, {Maruyama}, {Matsumoto}, {Miyaji}, {Nagayama}, {Nagayoshi},
  {Nakamura}, {Nakagawa}, {Namikawa}, {Omodaka}, {Oyama}, {Sakakibara},
  {Shimizu}, {Sora}, {Tsushima}, {Ueda}, {Ueda}, \& {Yamashita}}]{shintani08}
{Shintani}, M., {Imai}, H., {Ando}, K., {et~al.} 2008, \pasj, 60, 1077

\bibitem[{{Spencer} {et~al.}(1979){Spencer}, {Johnston}, {Moran}, {Reid}, \&
  {Walker}}]{spencer79}
{Spencer}, J.~H., {Johnston}, K.~J., {Moran}, J.~M., {Reid}, M.~J., \&
  {Walker}, R.~C. 1979, \apj, 230, 449

\bibitem[{{Sudou} {et~al.}(2019){Sudou}, {Omodaka}, {Murakami}, {Nagayama},
  {Nakagawa}, {Urago}, {Nagayama}, {Hirano}, \& {Honma}}]{sudou19}
{Sudou}, H., {Omodaka}, T., {Murakami}, K., {et~al.} 2019, \pasj, 71, 16

\bibitem[{{Sudou} {et~al.}(2017){Sudou}, {Shiga}, {Omodaka}, {Nakai}, {Ueda},
  \& {Takaba}}]{sudou17}
{Sudou}, H., {Shiga}, M., {Omodaka}, T., {et~al.} 2017, Journal of Korean
  Astronomical Society, 50, 157

\bibitem[{{Szymczak} \& {Engels}(1995)}]{szymczak95}
{Szymczak}, M. \& {Engels}, D. 1995, \aap, 296, 727

\bibitem[{{Szymczak} \& {Engels}(1997)}]{szymczak97}
{Szymczak}, M. \& {Engels}, D. 1997, \aap, 322, 159

\bibitem[{{Takaba} {et~al.}(2001){Takaba}, {Iwate}, {Miyaji}, \&
  {Deguchi}}]{takaba01}
{Takaba}, H., {Iwate}, T., {Miyaji}, T., \& {Deguchi}, S. 2001, \pasj, 53, 517

\bibitem[{{Takaba} {et~al.}(1994){Takaba}, {Ukita}, {Miyaji}, \&
  {Miyoshi}}]{takaba94}
{Takaba}, H., {Ukita}, N., {Miyaji}, T., \& {Miyoshi}, M. 1994, \pasj, 46, 629

\bibitem[{{Winnberg} {et~al.}(2008){Winnberg}, {Engels}, {Brand}, {Baldacci},
  \& {Walmsley}}]{winnberg08}
{Winnberg}, A., {Engels}, D., {Brand}, J., {Baldacci}, L., \& {Walmsley}, C.~M.
  2008, \aap, 482, 831 {\bf [Paper I]}

\bibitem[{{Yates} \& {Cohen}(1994)}]{yates94}
{Yates}, J.~A. \& {Cohen}, R.~J. 1994, \mnras, 270, 958

\bibitem[{{Zhang} {et~al.}(2017){Zhang}, {Zheng}, {Reid}, {Honma}, {Menten},
  {Brunthaler}, \& {Kim}}]{zhang17}
{Zhang}, B., {Zheng}, X., {Reid}, M.~J., {et~al.} 2017, \apj, 849, 99

\end{thebibliography}

\clearpage

\section{Appendix: all maser spectra}
In this section, we show all H$_2$O maser spectra of our targets. There are more spectra here than in the FVt-plots because in those plots (apart from having averaged spectra taken within four days of one another), we tried to avoid large gaps between observations in order to avoid having to interpolate the data over large time intervals.

\addtocounter{figure}{-1}

\begin{figure*}
\resizebox{18cm}{!}{
\includegraphics{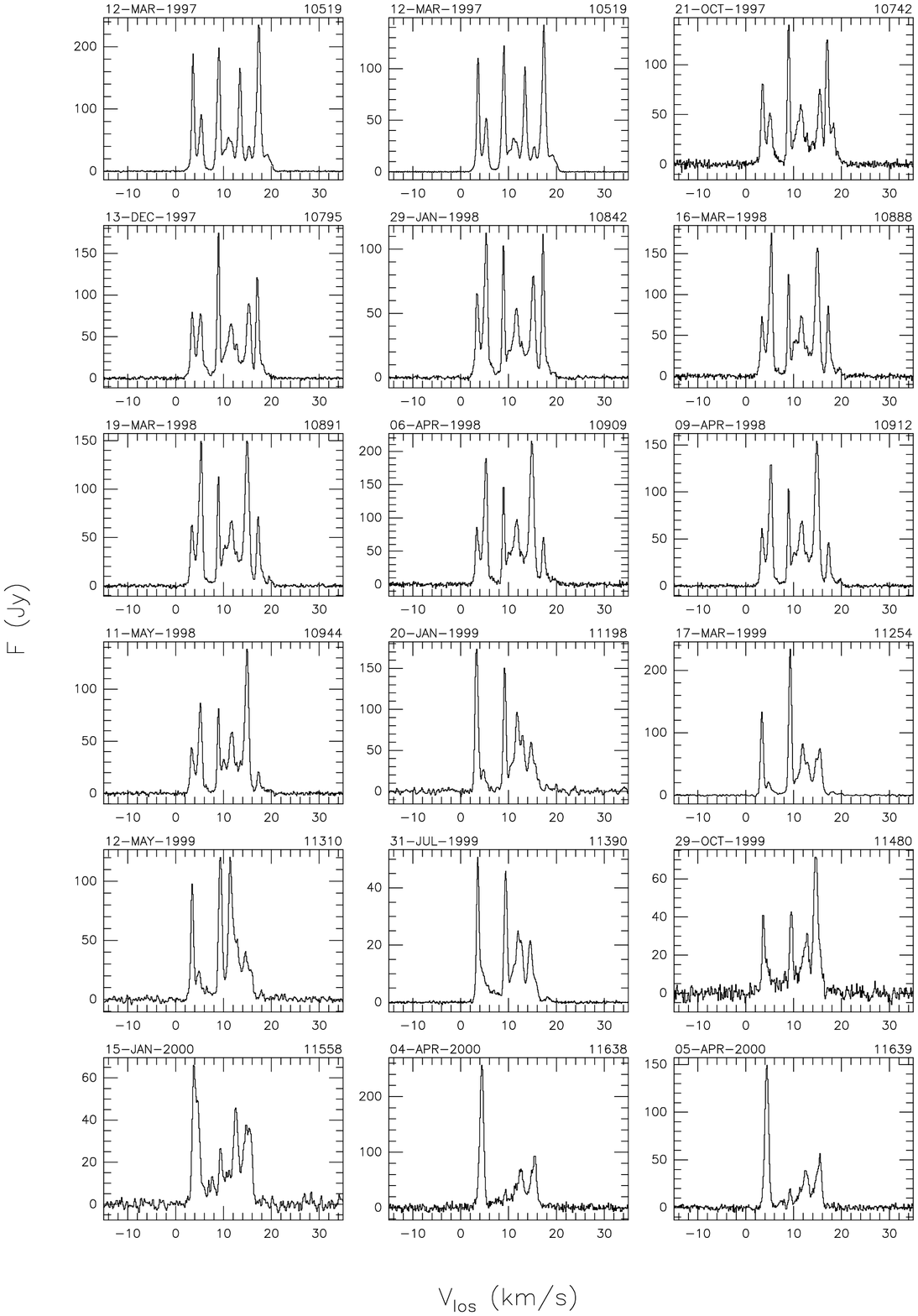}}
\caption{{\it R~Crt, continued.}}
\end{figure*}

\addtocounter{figure}{-1}

\begin{figure*}
\resizebox{18cm}{!}{
\includegraphics{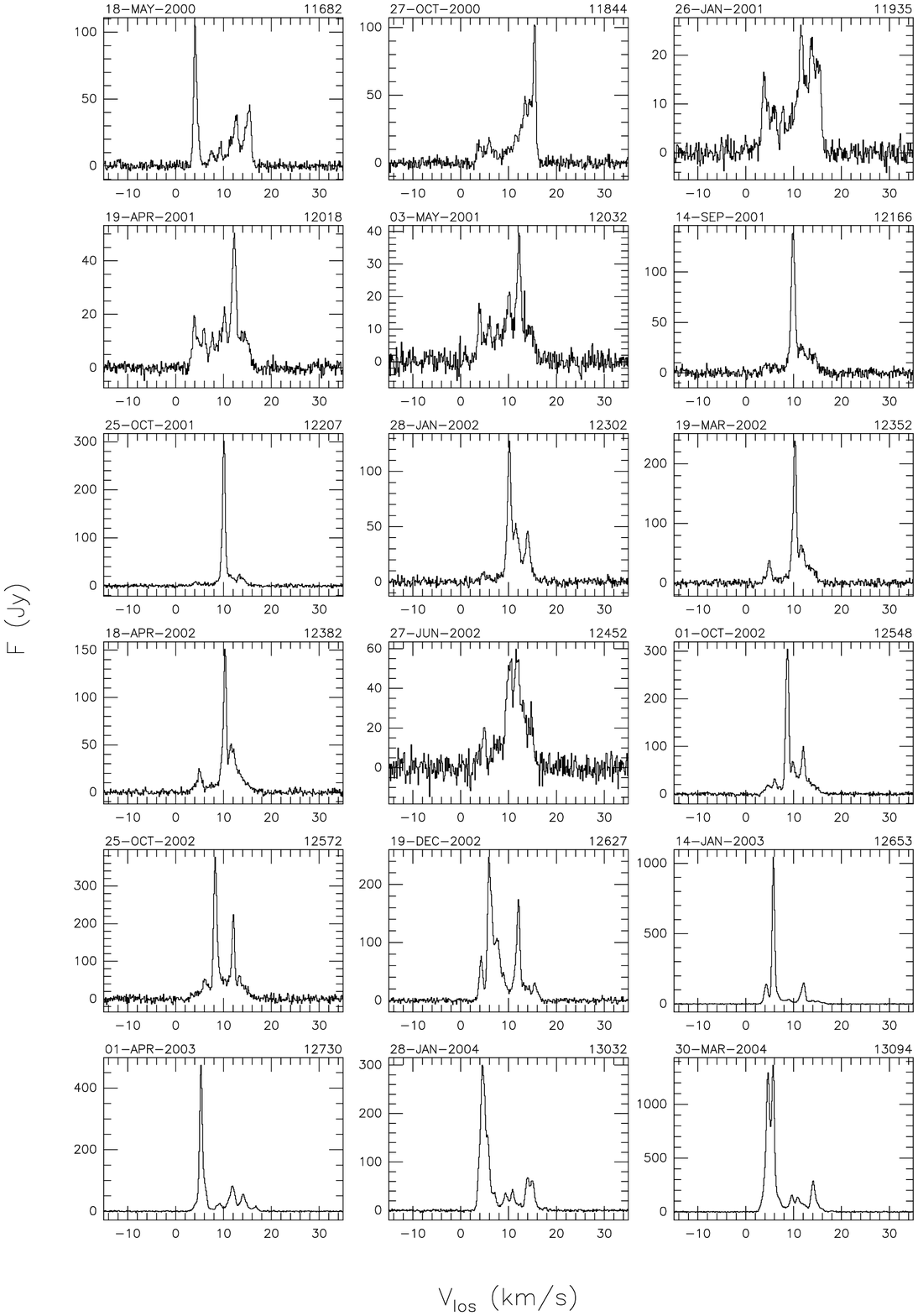}}
\caption{{\it R~Crt, continued.}}
\end{figure*}

\addtocounter{figure}{-1}

\begin{figure*}
\resizebox{18cm}{!}{
\includegraphics{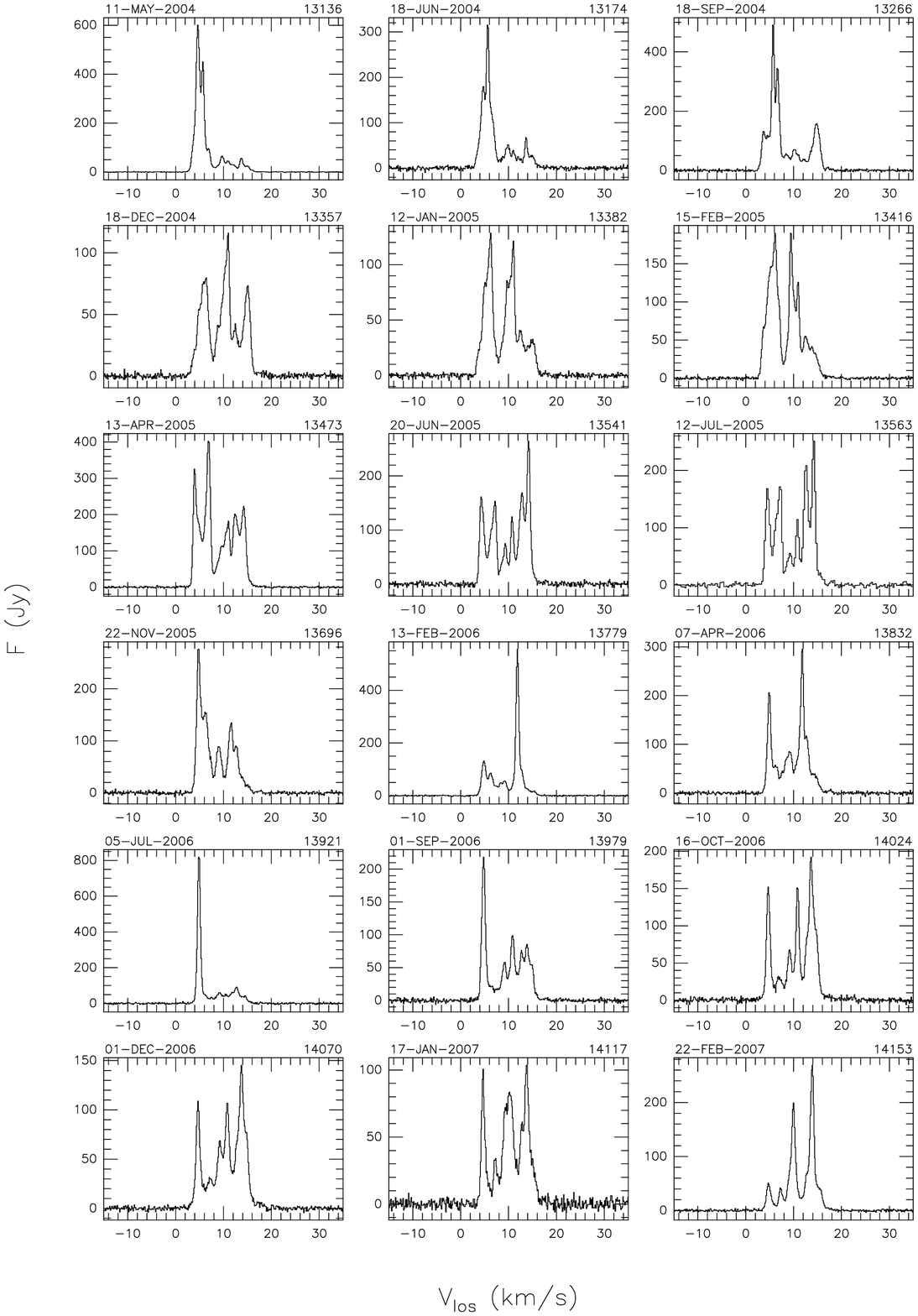}}
\caption{{\it R~Crt, continued.}}
\end{figure*}

\addtocounter{figure}{-1}

\begin{figure*}
\resizebox{18cm}{!}{
\includegraphics{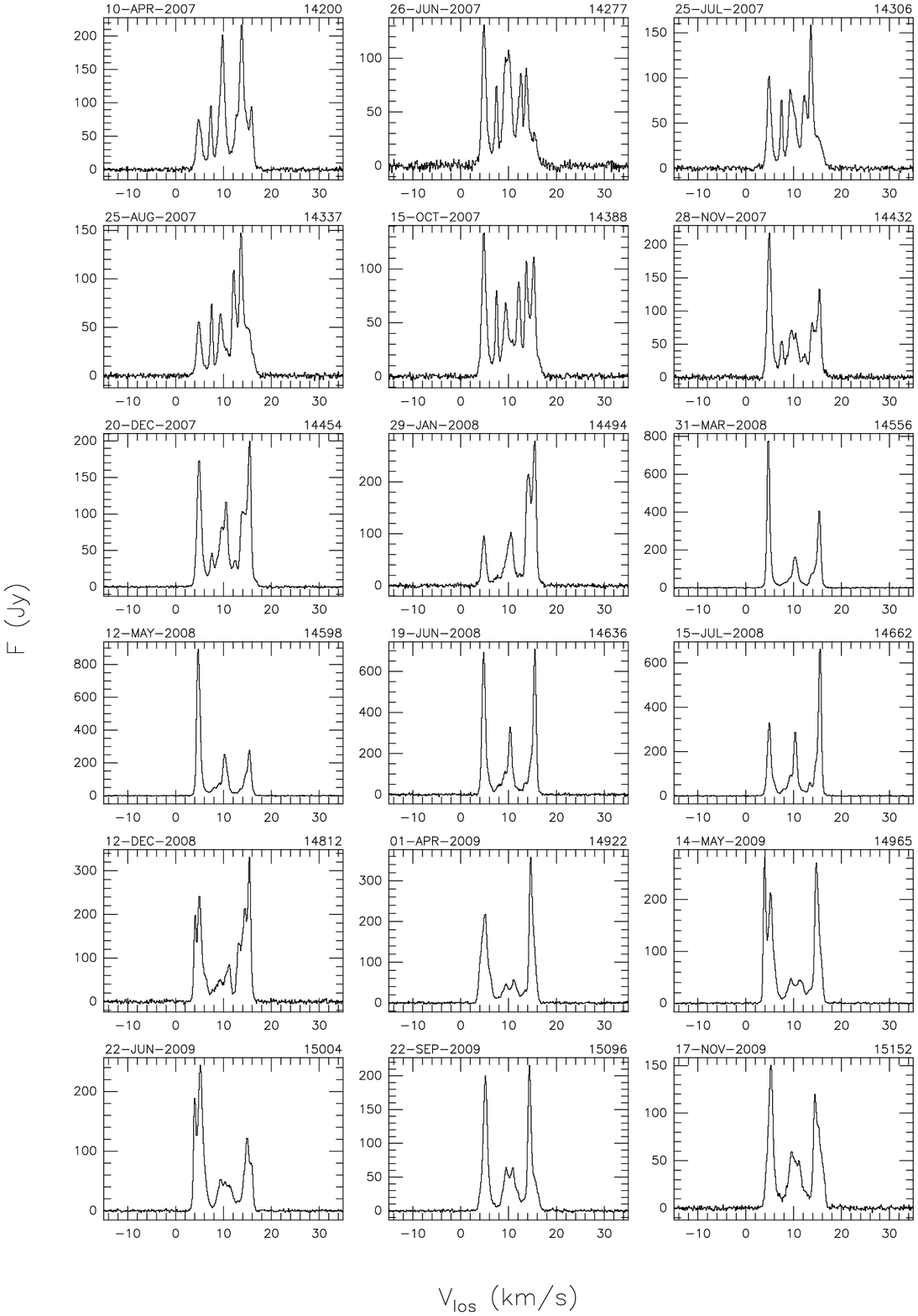}}
\caption{{\it R~Crt, continued.}}
\end{figure*}

\addtocounter{figure}{-1}

\begin{figure*}
\resizebox{18cm}{!}{
\includegraphics{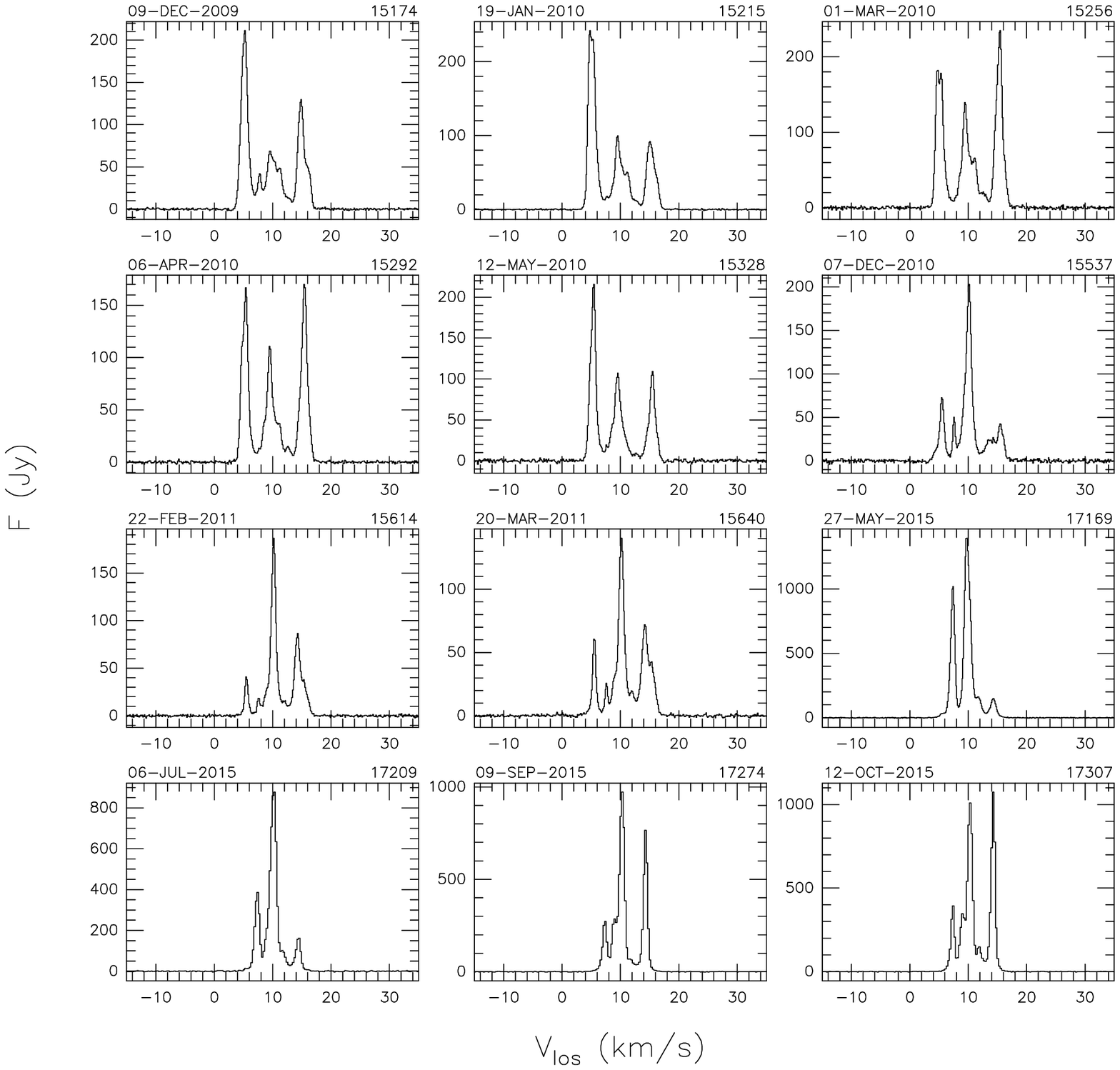}}
\caption{{\it R~Crt, continued.}}
\end{figure*}

\clearpage

\begin{figure*}
\resizebox{18cm}{!}{
\includegraphics{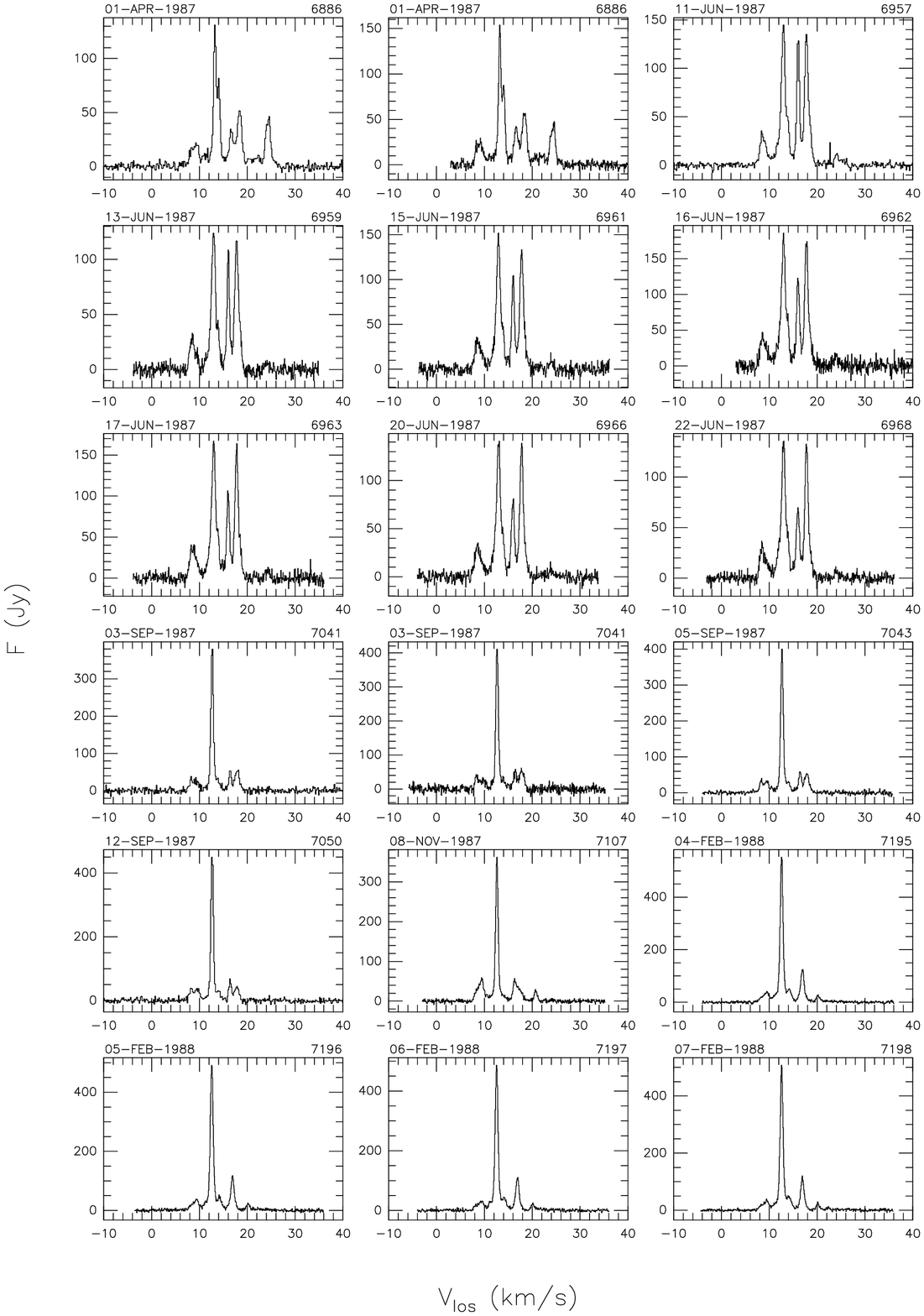}}
\caption{All H$_2$O maser spectra of RT~Vir. The observing date (top left) and TJD (top right) are indicated for each spectrum.}
\label{fig:rtvir_all}
\end{figure*}

\addtocounter{figure}{-1}

\begin{figure*}
\resizebox{18cm}{!}{
\includegraphics{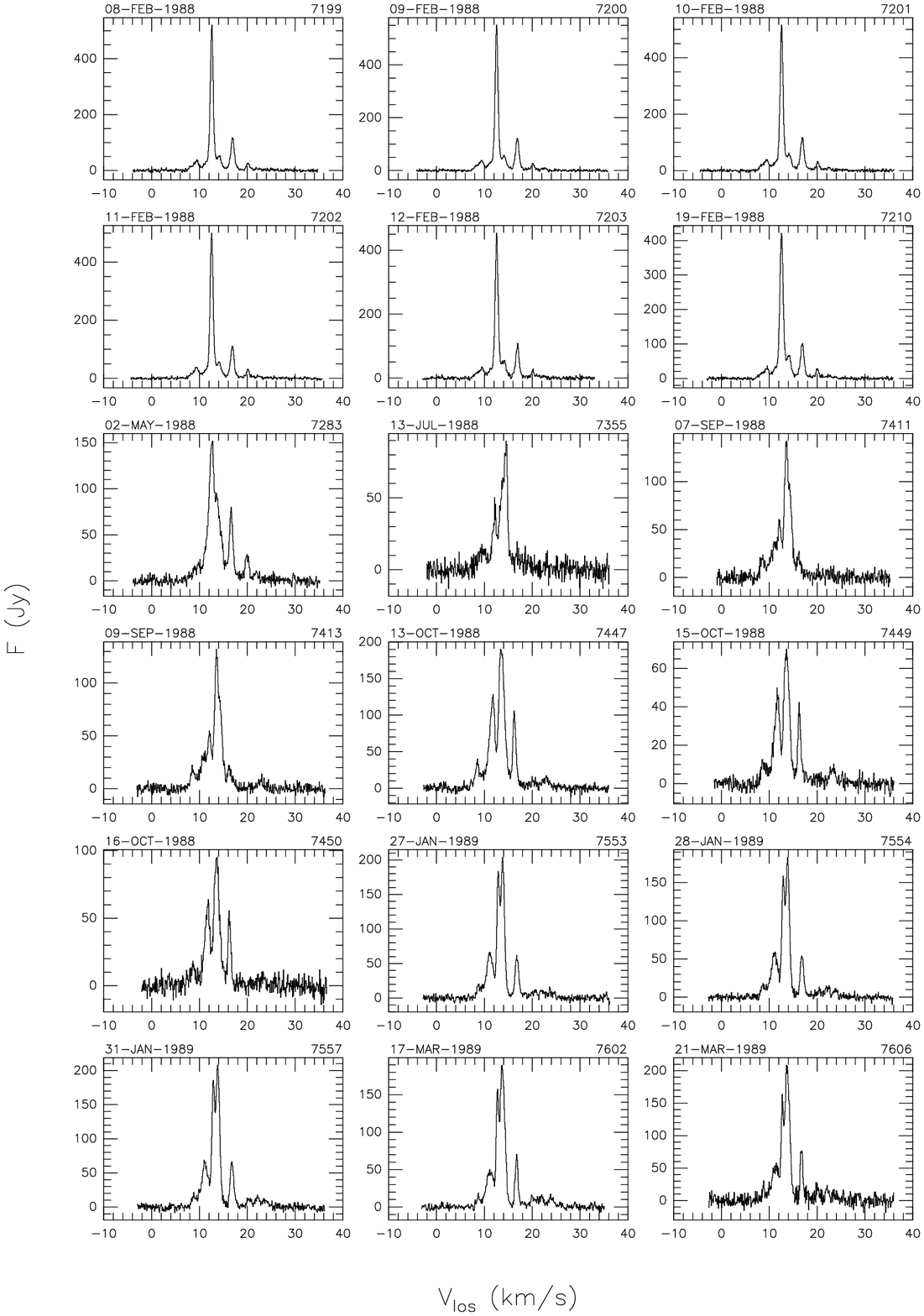}}
\caption{{\it RT~Vir, continued.}}
\end{figure*}

\addtocounter{figure}{-1}

\begin{figure*}
\resizebox{18cm}{!}{
\includegraphics{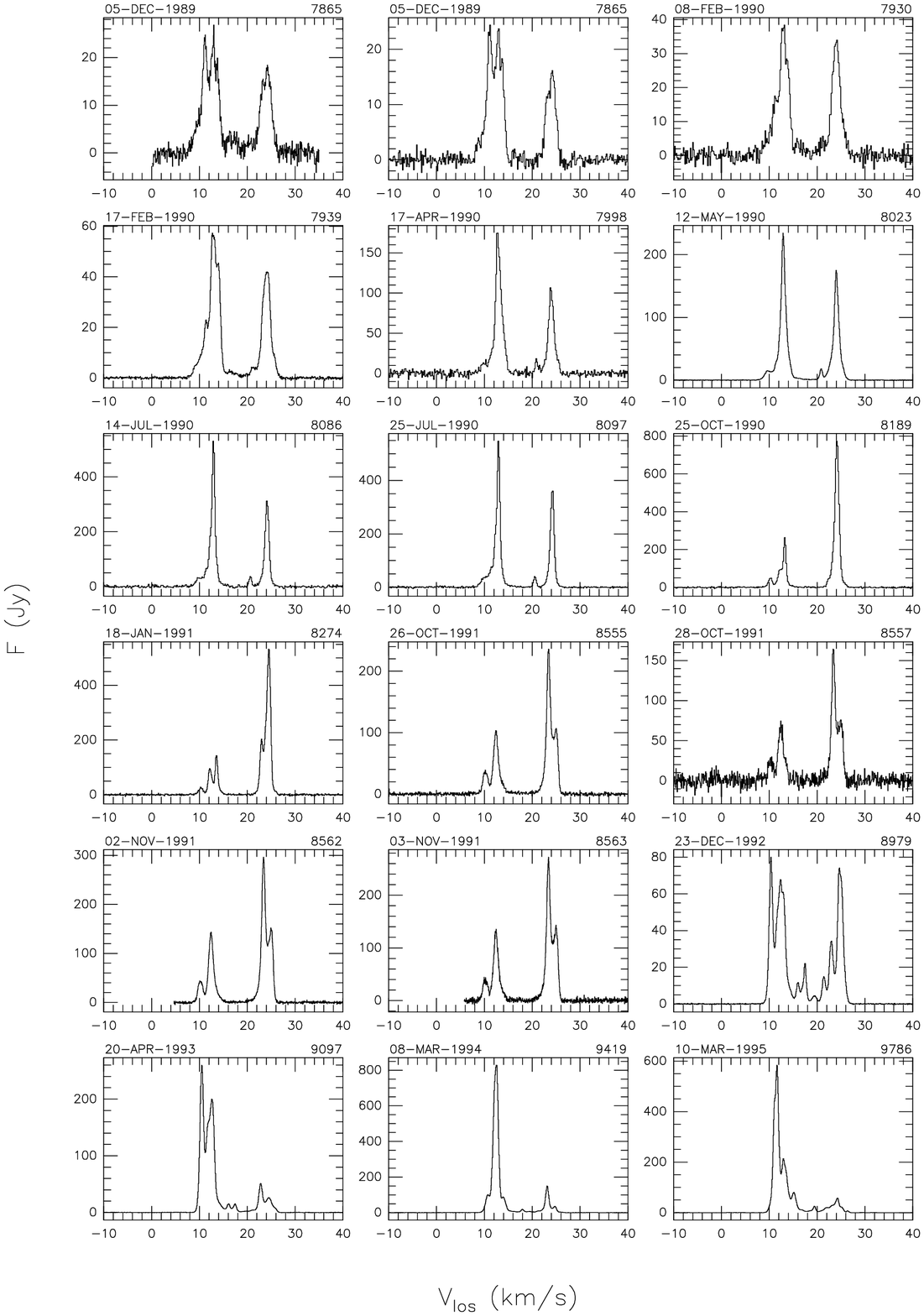}}
\caption{{\it RT~Vir, continued.}}
\end{figure*}

\addtocounter{figure}{-1}

\begin{figure*}
\resizebox{18cm}{!}{
\includegraphics{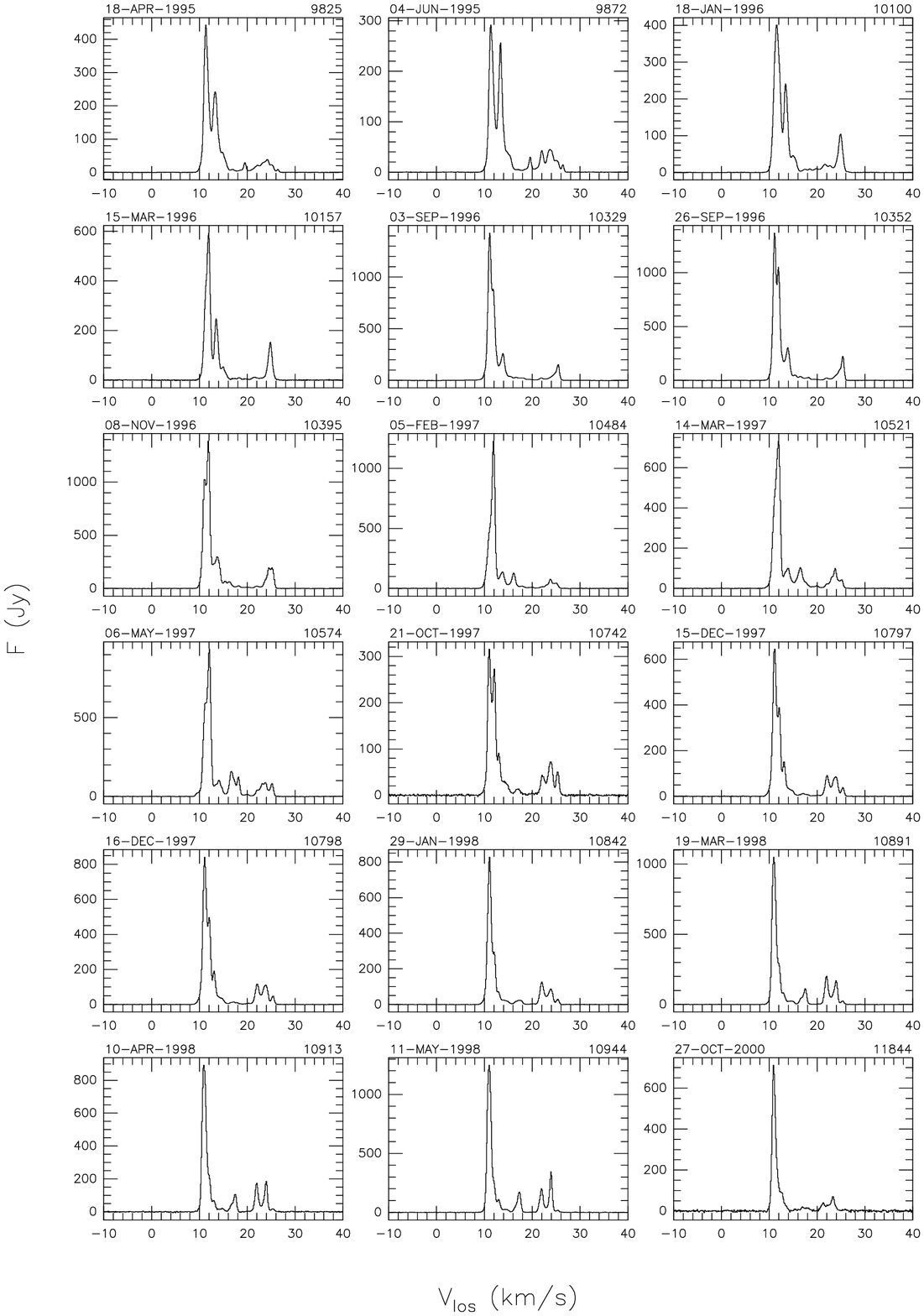}}
\caption{{\it RT~Vir, continued.}}
\end{figure*}

\addtocounter{figure}{-1}

\begin{figure*}
\resizebox{18cm}{!}{
\includegraphics{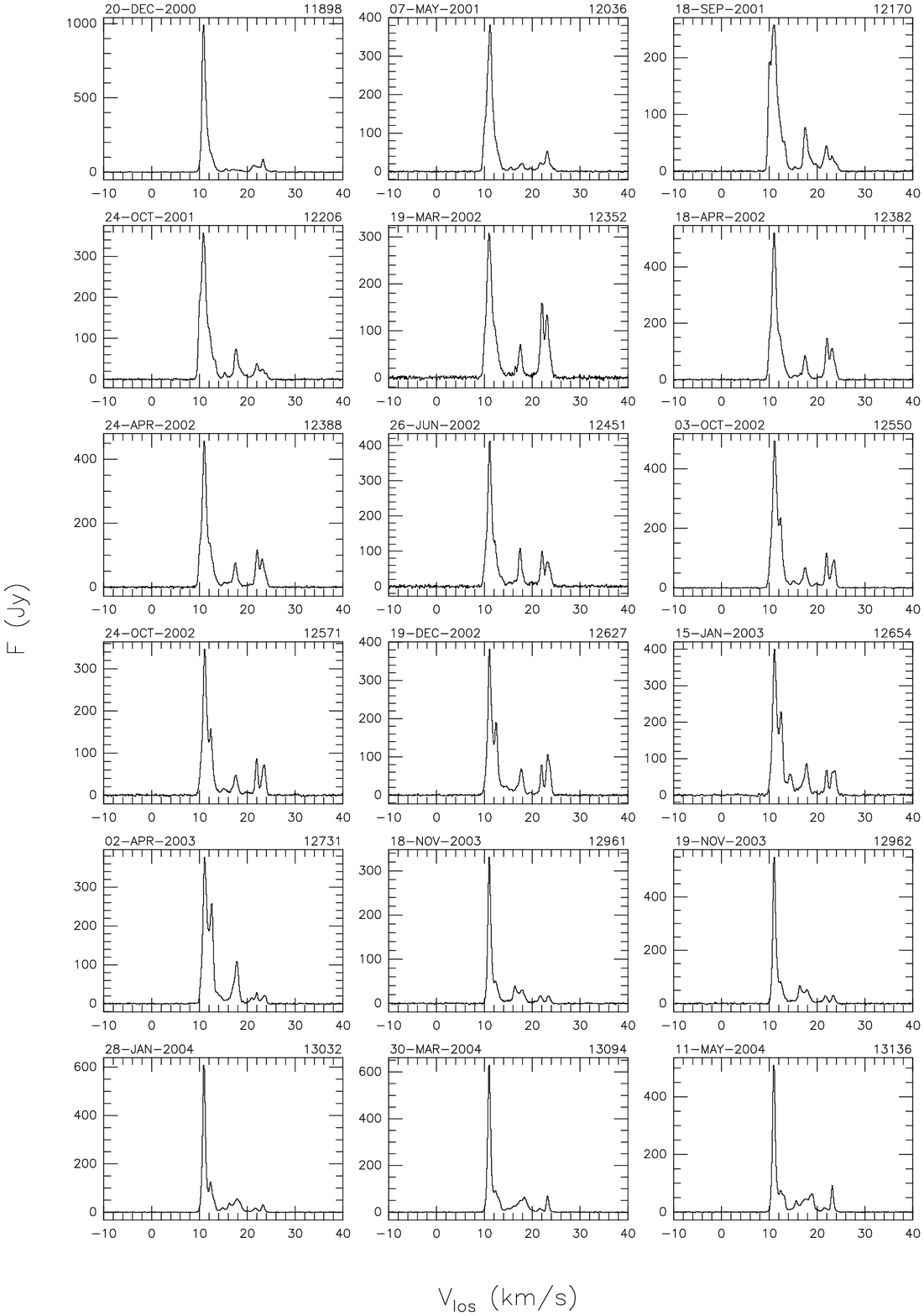}}
\caption{{\it RT~Vir, continued.}}
\end{figure*}

\addtocounter{figure}{-1}

\begin{figure*}
\resizebox{18cm}{!}{
\includegraphics{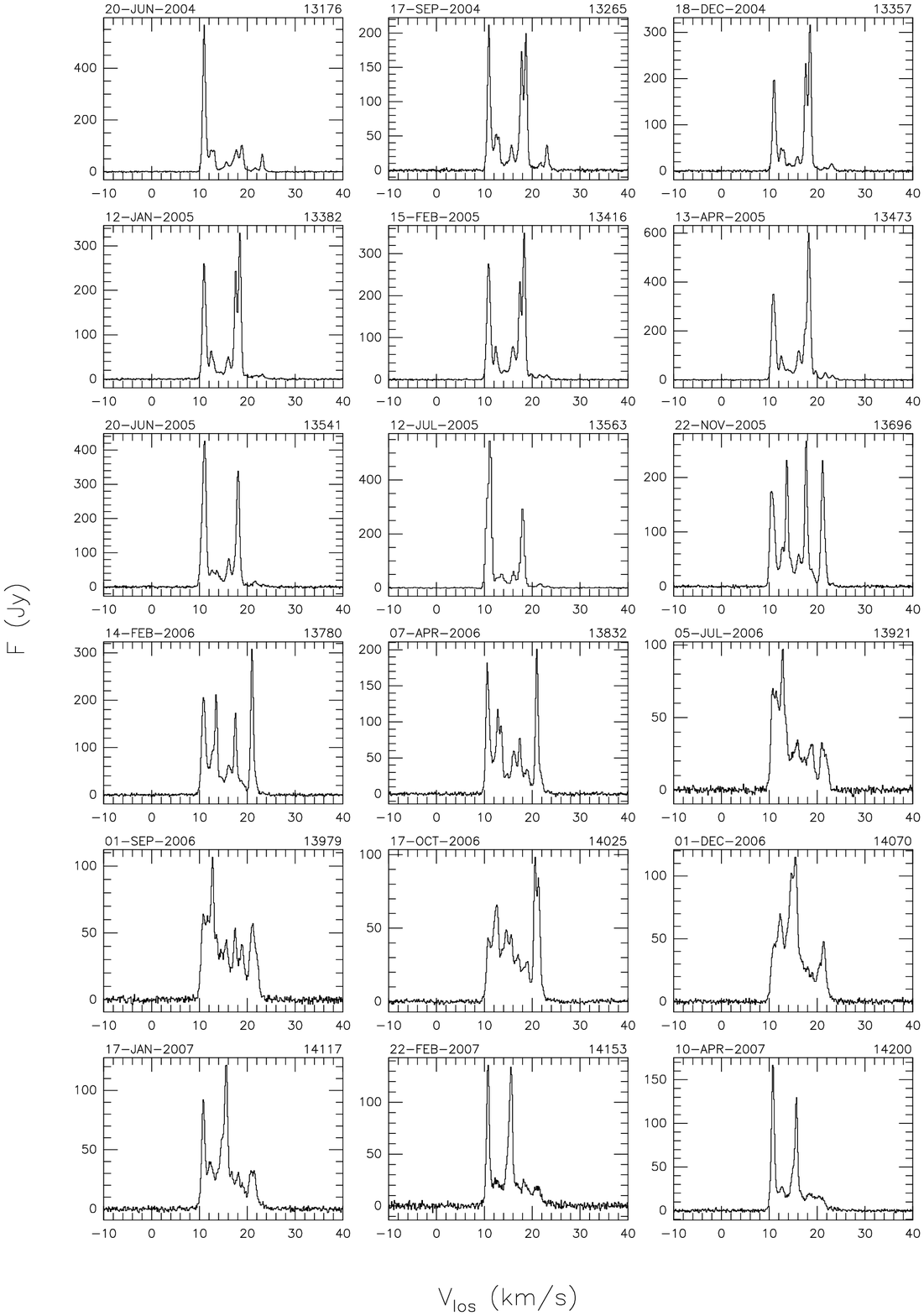}}
\caption{{\it RT~Vir, continued.}}
\end{figure*}

\addtocounter{figure}{-1}

\begin{figure*}
\resizebox{18cm}{!}{
\includegraphics{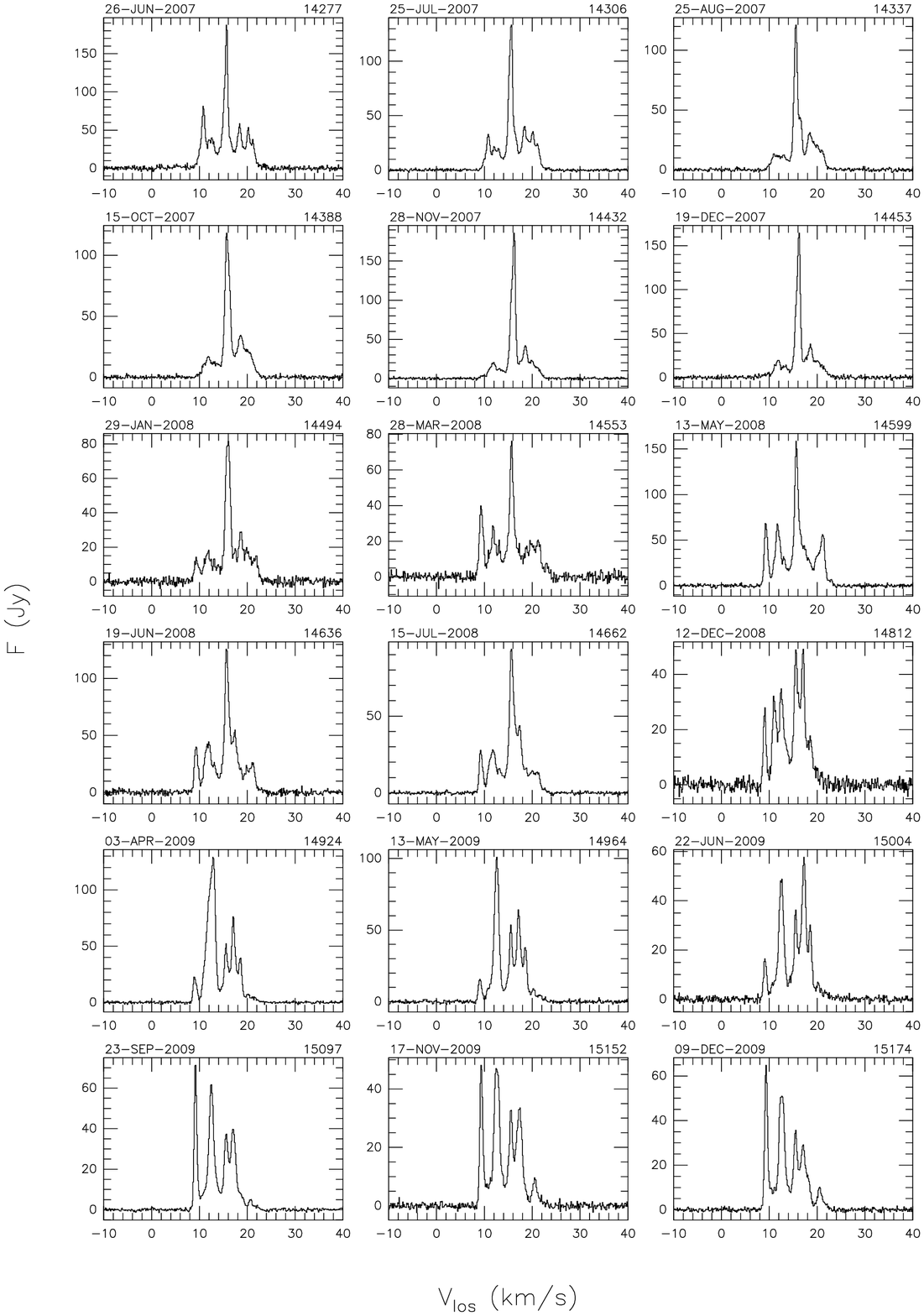}}
\caption{{\it RT~Vir, continued.}}
\end{figure*}

\addtocounter{figure}{-1}

\begin{figure*}
\resizebox{18cm}{!}{
\includegraphics{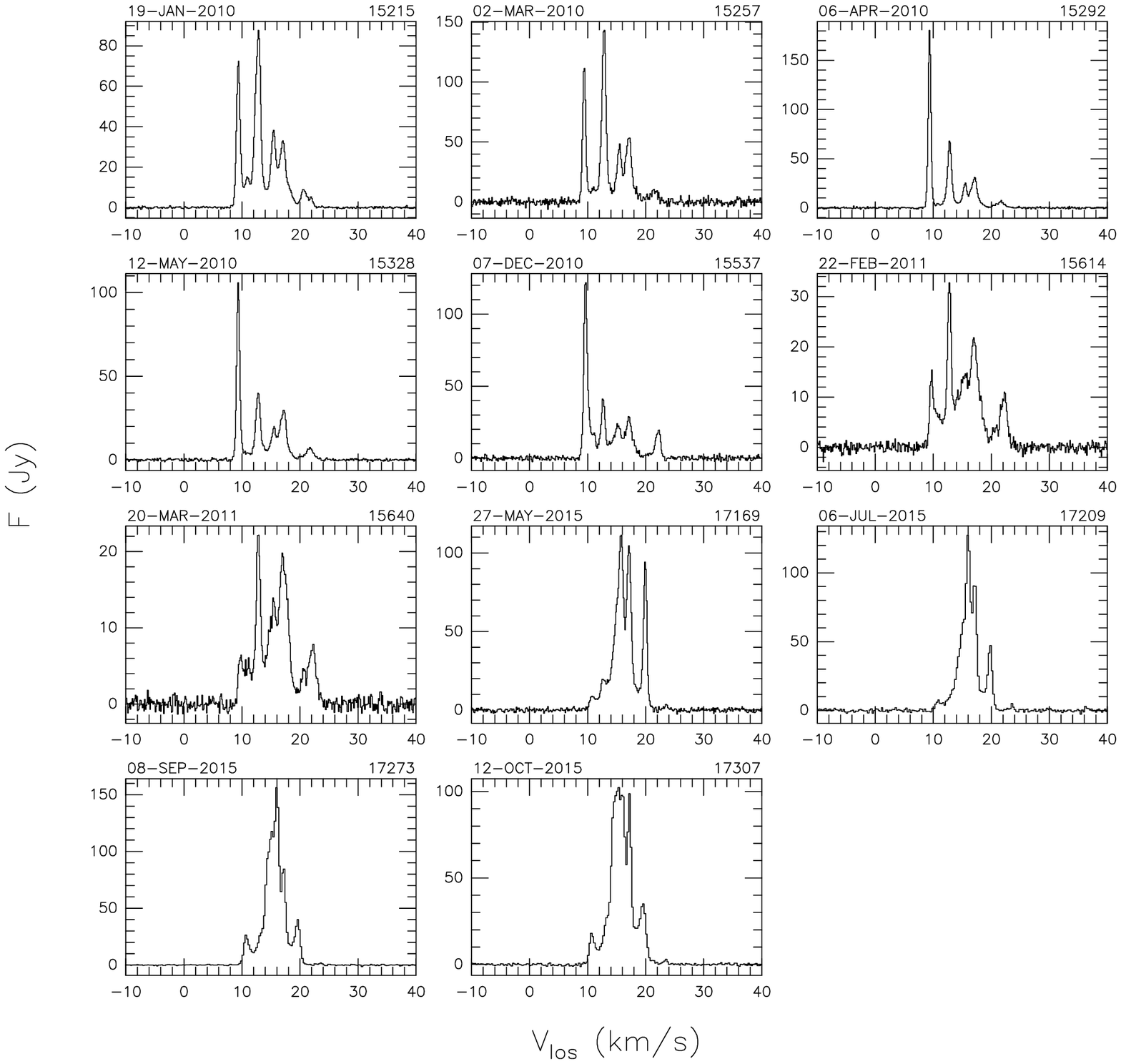}}
\caption{{\it RT~Vir, continued.}}
\end{figure*}

\end{document}